\documentclass{emulateapj}

\usepackage{psfig}
\usepackage{amsmath}
\usepackage{amssymb}
\usepackage{graphicx}
\usepackage{appendix}
\usepackage{color}
\usepackage{multirow}
\usepackage[normalem]{ulem}
\usepackage{xspace}
\usepackage{nicefrac}

\newcommand{\lSect}[1]{{\label{sec:#1}}}
\newcommand{\lFig}[1]{{\label{fig:#1}}}
\newcommand{\lEq}[1]{{\label{eq:#1}}}
\newcommand{\lTab}[1]{{\label{tab:#1}}}
\def\gtaprx {\lower .1ex\hbox{\rlap{\raise .6ex\hbox{\hskip .3ex
	{\ifmmode{\scriptscriptstyle >}\else
		{$\scriptscriptstyle >$}\fi}}}
	\kern -.4ex{\ifmmode{\scriptscriptstyle \sim}\else
		{$\scriptscriptstyle\sim$}\fi}}}
\def\ltaprx {\lower .1ex\hbox{\rlap{\raise .6ex\hbox{\hskip .3ex
	{\ifmmode{\scriptscriptstyle <}\else
		{$\scriptscriptstyle <$}\fi}}}
	\kern -.4ex{\ifmmode{\scriptscriptstyle \sim}\else
		{$\scriptscriptstyle\sim$}\fi}}}
\newcommand{\FIGFF}[2]{{\ref{fig:#2}{#1}}}

\newcommand{\FIG}[2]{{Fig.~\FIGFF{#1}{#2}}}
\newcommand{\Fig}[1]{{\FIG{}{#1}}}

\newcommand{\Sectff}[1]{{\ref{sec:#1}}}
\newcommand{\Sect}[1]{{\S\Sectff{#1}}}

\newcommand{\Eqalt}[1]{Eq.~\ref{eq:#1}}

\newcommand{\Msun}{\ensuremath{\mathrm{M}_\odot}}
\newcommand{\Lsun}{\ensuremath{\mathrm{L}_\odot}}
\newcommand{\Rsun}{\ensuremath{\mathrm{R}_\odot}}

\newcommand{\Tab}[1]{{Table~\ref{tab:#1}}}

\newcommand{\KEPLER}{\textsc{Kepler}\xspace}

\newcommand{\Mlo}{{\ensuremath{M_\mathrm{low}}}\xspace}
\newcommand{\Mhi}{{\ensuremath{M_\mathrm{high}}}\xspace}

\def\gtaprx {\lower .1ex\hbox{\rlap{\raise .6ex\hbox{\hskip .3ex
	{\ifmmode{\scriptscriptstyle >}\else
		{$\scriptscriptstyle >$}\fi}}}
	\kern -.4ex{\ifmmode{\scriptscriptstyle \sim}\else
		{$\scriptscriptstyle\sim$}\fi}}}
\def\ltaprx {\lower .1ex\hbox{\rlap{\raise .6ex\hbox{\hskip .3ex
	{\ifmmode{\scriptscriptstyle <}\else
		{$\scriptscriptstyle <$}\fi}}}
	\kern -.4ex{\ifmmode{\scriptscriptstyle \sim}\else
		{$\scriptscriptstyle\sim$}\fi}}}

\begin{document}


\title{The Pair-Instability Mass Gap for Black Holes}

\author{S. E. Woosley\altaffilmark{1} and Alexander Heger\altaffilmark{2-5}}
\altaffiltext{1}{Department of Astronomy and Astrophysics, University
  of California, Santa Cruz, CA 95064, woosley@ucolick.org}
\altaffiltext{2}{School of Physics and Astronomy, Monash University, Vic 3800, Australia}
\altaffiltext{3}{Australian Research Council Centre of Excellence for Gravitational Wave Discovery (OzGrav), Clayton, VIC 3800, Australia}
\altaffiltext{4}{Center of Excellence for Astrophysics in Three Dimensions (ASTRO-3D), Australia}
\altaffiltext{5}{Joint Institute for Nuclear Astrophysics, 1 Cyclotron Laboratory, National Superconducting Cyclotron Laboratory, Michigan State University, East Lansing, MI 48824-1321, USA}

\begin{abstract}
  Stellar evolution theory predicts a ``gap'' in the black hole birth
  function caused by the pair instability. Presupernova stars that
  have a core mass below some limiting value, \Mlo, after all
  pulsational activity is finished, collapse to black holes, while
  more massive ones, up to some limiting value, \Mhi, explode,
  promptly and completely, as pair-instability supernovae. Previous
  work has suggested $\Mlo \approx 50$\,\Msun\ and $\Mhi \approx
  130$\,\Msun. These calculations have been challenged by recent LIGO
  observations that show many black holes merging with individual
  masses, $\Mlo \gtaprx\, 65$\,\Msun.  Here we explore four factors
  affecting the theoretical estimates for the boundaries of this mass
  gap: nuclear reaction rates, evolution in detached binaries,
  rotation, and hyper-Eddington accretion after black hole
  birth. Current uncertainties in reaction rates by themselves allow
  \Mlo to rise to 64\,\Msun\ and \Mhi as large as 161\,\Msun. Rapid
  rotation could further increase \Mlo to $\sim70$\,\Msun, depending
  on the treatment of magnetic torques. Evolution in detached binaries
  and super-Eddington accretion can, with great uncertainty, increase
  \Mlo still further.  Dimensionless Kerr parameters close to unity
  are allowed for the more massive black holes produced in close
  binaries, though they are generally smaller.
 \end{abstract}


\section{INTRODUCTION}
\lSect{intro}

Pulsational pair-instability supernovae (PPISN) and pair-instability
supernovae (PISN) are of interest because of their potential roles in
early stellar nucleosynthesis \citep{Heg02}, in explaining unusual
types of supernovae \citep{Woo07}, and as the progenitors of massive
black hole systems that can merge and produce detectable gravitational
radiation signals \citep{Woo16}. Of particular current interest is the
prediction of a ``mass gap'' between about
50\,\Msun\ \citep[henceforth \Mlo;][]{Woo07,Bel16,Woo17,Woo19} and
130\,\Msun\ \citep[\Mhi;][]{Heg02} where very few black holes should
exist in close binaries that merge in a Hubble time.

The existence, or at least the boundaries of such a gap have been
challenged by the discovery of a single merging black hole pair in
which the more massive component had a mass near 85\,\Msun\ \citep{Abb20a,Abb20b}, and, more recently, by the publication of a
catalog of sources \citep{Abb20c} in which numerous black holes
merging by gravitational radiation have masses in the 60 to 70\,\Msun
\ range. Other recent theoretical papers have pointed out the
sensitivity of the boundaries of the putative mass gap to
uncertainties in reaction rates \citep{Far20,Cos20}, the effects of
rotation \citep{Mar20}, and super-Eddington accretion \citep{Van20}.
These papers were submitted while the present one was in preparation,
and now take precedence. It remains, useful, however to examine the role
of this physics using different sets of assumptions in a single code and to
attempt an evaluation of which of the causes is dominant.

Specifically we consider four possibilities for producing black hole
remnants in close binaries with individual masses substantially
heavier than 50\,\Msun: \textsl{1)} A decrease in the rate for the
$^{12}$C($\alpha,\gamma)^{16}$O and/or an increase in the rate for the
triple-alpha ($3\alpha$) reaction (\Sect{carbon}). The range of rates
considered is only what is currently favored by experiment. Our
choices might thus be regarded as conservative, but are influenced
both by experiment and nucleosynthetic considerations
(\Sect{nucleo}). These changes alone can potentially raise \Mlo to
$64$\,\Msun, but are unlikely, by themselves, to increase \Mlo to
85\,\Msun. \textsl{2)} Making the black hole in a detached binary
(\Sect{single}). Using the variation in rates discussed in
\Sect{carbon}, it is possible to produce black holes of 100\,\Msun\ or
more if the star's hydrogen envelope is retained at the time of core
collapse. The obvious problems here are the low mass loss rate
required and the difficulty getting the two black holes together at
the end, if one of them was made in a supernova that was a giant star
when it exploded. \textsl{3)} The effects of rapid rotation
(\Sect{rotation}). The dynamical effects of rotation alone can only
increase \Mlo modestly \citep{Mar20}, but rotation coupled with
uncertain reaction rates can raise \Mlo to $\sim70$\,\Msun. In the
most extreme cases, the angular momentum in the presupernova star is
too great to form a black hole without the loss of angular momentum
and an uncertain amount of mass. \textsl{4)} Super-Eddington accretion
after the first black hole has formed (\Sect{accrete}). As the
secondary expands just following hydrogen depletion, a stable phase of
frictionally-driven mass loss \citep{Pod01} might substantially increase
the mass of the black hole produced by the primary. This involves some
very uncertain physics, but is the most likely way to produce
black holes well over 70\,\Msun\ in a single close binary.

Our computational approach is the same as in \citet{Woo17}, except
with the addition of centrifugal force to the structure equations
(\Sect{rotation}). All models are calculated using the \KEPLER code
\citep{Wea78,Woo02} and the physics described in those papers.  In
particular, the previous standard values for the reaction rates that
are varied are taken from \citet{Deb17} and \citet{Buc96a} for
$^{12}$C($\alpha,\gamma)^{16}$O and \citet{Cau88} for $3
\alpha$. Unless otherwise noted, mass loss rates for giant stars,
which are relevant for \Sect{single}, are taken from \citet{Nie90} and
\citet{Yoo17}, the latter being an amalgamation of \citet{Han14} and
\citet{Tra16}.

\section{Variable Carbon Production}
\lSect{carbon}

The pair-instability begins mildly, following the central
depletion of helium. As the star contracts towards carbon and,
ultimately, oxygen ignition, more pairs are produced at the higher
temperature required for burning and the instability becomes
stronger. Central carbon burning is net endoergic in stars of this large
mass, with the energy generation rate never exceeding neutrino
losses, and hence proceeding in a radiative fashion.  Far enough out, however, carbon shell burning becomes exoergic, even in the face of
neutrino losses, and can power convective shells whose strength and
extent affects the subsequent dynamics of the implosion. Helium shell
burning in the outer layers of the star can have a similar, though
usually smaller effect.

After central carbon depletion, infalling matter gathers momentum that
often, but not always results in dynamical overshoot and explosive
oxygen burning. A variable carbon abundance in the outer layers of the
star can provide varying degrees of support as the core becomes
unstable. This can dramatically influence the strength of the
instability and the energy, or even the possibility of a subsequent
explosion \citep{Far20}. Basically a strong carbon burning shell
provides an energy source that increases as the star
contracts. Sufficient off-center burning results in oxygen burning
igniting stably rather that catastrophically, and the subsequent
evolution is very different. Oxygen burning in a shell can still
become unstable later, even during silicon burning, but there is less
fuel to burn then, and the instability usually results in less matter being ejected.

\subsection{Reaction Rates}
\lSect{rates}

Substantial adjustments to the carbon abundance after helium burning
are possible because of uncertainties in the critical reaction rates
for $3\alpha$ and $^{12}$C($\alpha,\gamma)^{16}$O and the treatment of
convective mixing \citep{Far19,Far20,Ren20}.  Historically, the rate for
$^{12}$C($\alpha,\gamma)^{16}$O during helium burning has been a major
source of uncertainty for studies of stellar evolution and
nucleosynthesis \citep{Wea93,Tur10,Wes13}. This rate is characterized by an
``$S$-factor'' evaluated at a typical Gamow energy for reactions during
helium burning, which is taken to be 300 keV. \citet{Deb17} have
reviewed the experimental situation and offered their own analysis,
concluding that this critical $S$-factor is most likely $140 \pm 21
\ ^{+18}_{-11}$\,keV\,b. This is smaller than some other recent
publications (see their Table IV) which gave a value near 165 keV b,
but usually with larger error bars. In particular, \citet{Buc96a},
which serves as a reference point in the present study, gave a best
value of $S_{300} = 146$\,keV\,b with a possible range from 62 to 270\,kev,b \citep[see also][]{Buc06}. In the same time frame \citet{Buc96b}
found a best value of $S_{300}$ =$165 \pm 75$\,keV\,b. Here we shall
consider a possible range of $S$-factors from 110 to 205 keV b with a
standard value 146\,keV\,b.  The lower bound is that of
\citet{Deb17}. The upper bound is a bit larger than theirs, and more
consistent with \citet{Buc96b}. Past studies of nucleosynthesis
(\Sect{nucleo}) have favored such large values. Computationally these
variations were effected by multiplying the standard
fitted rate of \citet{Buc96a} by factors, $f_{\rm Buch}$, ranging from
0.75 to 1.4 (\Tab{helium}). As will be shown, \Mlo is more
sensitive to the rate for $^{12}$C($\alpha,\gamma)^{16}$O when it is
small than when it is large, which justifies, for now, not exploring
larger multipliers than 1.4.

\citet{She20} have given a recent reevaluation
of the electric quadrupole part of the reaction rate $S_{\rm E2}$ = 80
$\pm 40$\,keV\,b which is larger than the 45\,keV\,b adopted by
\citet{Deb17} and more consistent with \citet{Buc96b}. They recommend
that the error bar on the \citeauthor{Deb17} rate be raised from +18 to +39\,keV\,b, i.e., the total $S$-factor is more likely to be larger than 146\,keV\,b than smaller.


The rate for the $3 \alpha$ reaction depends on the radiative width
for the ``Hoyle state'', a $0^+$ resonance at an excitation energy of
7.65\,MeV in $^{12}$C. A small correction is generally added to the
radiative width for pair emission. Summarizing previous work,
\citet{Fre14} gave what might be considered the previous standard value
for this width,
$\Gamma_{\rm rad} = 3.78$\,meV (i.e., $3.78 \times 10^{-3}$\,eV). This is
close to the value used in \citet{Cau88}, $\Gamma_{\rm rad+pair} =
3.70$\,meV \citep[Hale in][]{Wal97}.  New recent measurements
\citep{Kib20,Eri20}, though by no means the final word, give
$\Gamma_{\rm rad} = 5.1(6)$\,meV, implying an upward revision of about
$35\,\%$. This large increase is substantially outside the error bar
of previous measurements and has major implications for stellar
nucleosynthesis (\Sect{nucleo}), but is considered here, for now, as a
limiting case.  Traditionally this reaction rate was assumed to be
known to $15\,\%$ accuracy \citep[e.g.,][]{Rol88}. In the present
study, we define $f_{3\alpha}$ as the multiplier on the \citet{Cau88}
rate for $3 \alpha$. This is $f_{3\alpha}$ = 1.35 corresponds
approximately to the new rate of \citet{Kib20}.


\subsection{Nucleosynthetic limits}
\lSect{nucleo}

In our past works, including \citet{Woo17}, the multiplying factor in
\KEPLER for the \citet{Buc96a} recommended value (146\,keV\,b) was
$f_{\rm Buch}$ = 1.2. Within the context of a then very large range of
possible values for the $S$-factor, \citet{Wea93} carried out a survey
of nucleosynthesis in massive stars to determine what value worked
best for nucleosynthesis, finding $S_{300} = 170 \pm\,20$\,keV\,b.
That value was used in \KEPLER prior to 1996. After 1996, this was
refined to be 1.2 times the rate given by \citet{Buc96a}, or $S_{300}= 175$\,keV\,b. This value has remained constant ever since, in part, to
facilitate comparison among various calculations carried out with the
\KEPLER code, but has always remained within the experimental error
bar and close to its centroid.

\citet{Tur07} and \citet{Tur10} studied the influence of uncertainties
in the $3\alpha$ and $^{12}$C($\alpha,\gamma)^{16}$O reactions on
nucleosynthesis and found that an increase in the $3 \alpha$ rate of
10\,\% had about the same effect as an 8\,\% decrease in
$^{12}$C($\alpha,\gamma)^{16}$O. Hence, for small variations, it is
the the ratio of the reaction rates,
$\lambda_{\alpha\gamma}(^{12}C)/\lambda_{3\alpha}$ (or
$f_{3\alpha}$/$f_{\rm Buch}$), that matters most.  \citet{Wes13} and
\citet{Aus14} considered constraints on the ratio of these two rates
required to fit nucleosynthesis in massive stars.  They found that
multipliers that gave a best fit obeyed the relation $f_{\rm Buch}$ =
1.0 $f_{3 \alpha} +(0.35 \pm 0.2)$, including the effect on the
s-process (their $R_{\alpha,12}$ and $R_{3 \alpha}$ are equivalent to
our $f_{\rm Buch}$ and $f_{3\alpha}$). The best fit using only the
intermediate-mass isotopes was $f_{\rm Buch} = 1.0 f_{3\alpha} +
(0.25\pm 0.3)$. That is, nucleosynthesis favored a relatively large
rate for $^{12}$C($\alpha,\gamma)^{16}$O, $f_{\rm Buch} \sim 1.3$, and
any increase in $3\alpha$ required a commensurate increase in the rate
for $^{12}$C($\alpha,\gamma)^{16}$O.

We will be exploring values outside this recommended range,
including values of $f_{\rm Buch}$ as low as 0.75 and $f_{3\alpha}$ as large as
1.35. This large decrease in the ratio of the rates is necessary to
appreciably raise the black hole mass, but may have negative
implications for nucleosynthesis that will need to be explored in the
future. See Fig.~4 of \citet{Wes13}.

\subsection{Non-rotating Helium Stars of Constant Mass}
\lSect{helium}

In a very close binary, where little or no extended hydrogen envelope
can be tolerated, the evolution of pure helium stars can be used to
approximate the final presupernova state
\citep[e.g.,][]{Woo17}. Helium stars or carbon-oxygen stars will also
be the end state of evolution where the metallicity is sufficiently
high that the envelope is lost to a wind or the rotation so rapid as
to induce chemically homogeneous evolution \citep{Mae87} on the main
sequence. Once revealed the helium star will also experience mass loss
by winds, especially if the metallicity is large. During the
post-helium burning evolution, the evolution is so fast that
mass loss by winds is negligible. As will be shown (\Sect{mdot}),
helium cores evolved at constant mass are a good approximation to the
evolution of heavier helium cores that include mass loss. It is the
final mass that matters.

This simplification motivates a study of reaction-rate sensitivity in
helium stars stars of constant mass. Similar studies have been carried
out by \citet{Far19} and \citet{Far20}. We consider here the variation
of just the rates for $^{12}$C($\alpha,\gamma)^{16}$O and 3$\alpha$
because of their key role in determining the carbon abundance when the
star becomes unstable. It is assumed that once an iron core is formed,
typically around 2\,\Msun, that iron core and all the remaining star
collapses to a black hole. This will not necessarily be the case if
the star is very rapidly rotating (\Sect{rotation}).

Models are calculated using the \KEPLER code and the same physics and
zoning as in \citet{Woo17}. Convection is left on during interpulse
periods, but only in the core where where hydrostatic equilibrium
prevails. Convection speeds are always limited to $20\,\%$ sonic, and
convective velocities do not adjust instantaneously, but with a growth
rate that is limited to the (inverse of the) convective time scale.
Interpulse convection can sometime play an important role in mixing
combustible fuel, i.e., helium and carbon to deeper depths in the star
where it affects the strength of the next pulse. All models contained
in excess of 2,400 mass shells, with finer zoning near the center and
surface. Typical simulations took about 20,000 time steps, though
sometimes much more, to reach core collapse. Nuclear energy generation
was generally treated using the 19 isotope approximation network in
\KEPLER and the $128$ isotope quasiequilibrium network once the
central oxygen mass fraction fell below $0.05$. About 20 models were
also calculated using an adaptable network \citep{Rau02} with
typically 250 isotopes from hydrogen to germanium coupled directly to
the structure calculation. Changing networks caused final remnant
masses to vary by at most 1\,\Msun\ except near the critical mass where
the star fully exploded after the first pulse (\Sect{mhi}). Small
differences made a large difference in outcome for these cases, but
the value of \Mlo for a given choice of nuclear physics was not
greatly affected.

\subsubsection{Lower Boundary to the Pair-Instability Mass Gap}
\lSect{mlow}

Results for \Mlo are given in \Tab{helium}. As discussed by
\citet{Far19}, decreasing the rate for
$^{12}$C($\alpha,\gamma)^{16}$O, or raising the rate for 3$\alpha$
results in a larger amount of carbon being present in the outer part
of the star when the pair-instability is first encountered after
central carbon depletion.  The energy generated in these shells
inhibits the collapse to oxygen ignition and thus makes the explosion
milder. In cases with large carbon abundance, i.e., low
$^{12}$C($\alpha,\gamma)^{16}$O or high $3\alpha$, the energy from
carbon shell burning can even trigger its own series of weak
pair-instability pulsations. This affects the final remnant mass very
little, but might provide a low mass shell around the star when it
finally dies.

A key point is that if the carbon abundance is large enough, central
oxygen burning can ignite stably with energy transported by
convection. If a substantial amount of oxygen burns stably, then the
reservoir of fuel for future explosive events is reduced, even if they
occur. The pulsational pair instability is thus weakened. Less matter
is ejected with lower energy. This trend of weaker pulsations when
the carbon abundance is high is apparent in \Tab{helium}. Though the
focus here is on remnant masses, this behavior also has
important implications for the properties of any supernovae resulting
from the pulsational pair instability. The energies are still
sufficient to power a bright Type II supernova in stars where a
hydrogen envelope is retained. The light curves of these supernova
will lack radioactive tails and eject almost no new heavy elements,
but the energy and brightness of supernovae resulting from colliding
shells will be less. For unfavorable values of reaction rates, this
makes PPISN even less likely to produce superluminous supernovae
\citep{Woo17}.

For models with small carbon abundances, i.e., large
$^{12}$C($\alpha,\gamma)^{16}$O or small $3\alpha$ rates, the
characteristics of PPISN are not very sensitive to the rates. Focusing
on just those models where only $^{12}$C($\alpha,\gamma)^{16}$O is
varied, the maximum black hole mass is 46, 46, 48, and 56\,\Msun\ for
rate multipliers $f_{\rm Buch}$ = 1.4, 1.2, 1.0, and 0.8. For the
larger multipliers, the carbon burning shells are weak and the pair
instability is strong. These were the cases explored by
\citet{Woo17}. For central carbon mass fractions at helium burning
greater than about 0.15 though, the nature of the solution changes
dramatically. For the most extreme cases considered here, the pair
instability is essentially absent.  Only some weak pulses occur near
the end of the stars life when the core is already well into silicon
burning.  For intermediate values, the pair instability is still
present, but suppressed to a varying extent.

Increasing the $3 \alpha$ rate can further raise \Mlo.
For $f_{\rm Buch} = 0.75$ and $f_{3 \alpha} = 1.35$, the maximum black
hole mass is near 64\,\Msun. Larger values could be obtained by further
manipulation of the rates \citep{Far19,Far20}, and by including rotation
(\Sect{rotation}), but pending further experimental studies, we regard
such variations of reaction rates as unlikely, especially given the
potential problems with nucleosynthesis.

\subsubsection{Upper Boundary to the Pair-Instability Mass Gap}
\lSect{mhi}

Uncertainties in the reaction rates can also have an important effect on the
upper boundary of the pair-instability gap (\Tab{pisn}).  The least
massive black hole above the gap, \Mhi, now ranges from 136 to
161\,\Msun\ for the same choices of reaction rates as in
\Tab{helium}. There is a notable shift for $f_{\rm Buch} = 1.2$
from the value given by \citet{Heg02}, $\Mhi=133.3$\,\Msun, to
the new value, $\Mhi= 139$\,\Msun, a change of 4.3\,\% resulting
solely from different code physics.

About half of this change comes from directly coupling a large
reaction network of about 200 isotopes to the stellar structure
calculation, taking smaller time steps, and other minor changes to the
\KEPLER code in the last 20 years. The calculations of \citet{Heg02}
only used a 19 isotope approximation network for energy generation and
``coprocessed'' with a much larger network to obtain
nucleosynthesis. The quasiequilibrium network in \KEPLER was not used
due to problems at that time with convectively coupled zones containing
a varying amount of oxygen at explosive temperatures.

The other half comes from different assumptions regarding convection
during the implosion.  In \citet{Heg02}, it was assumed that
convection did not occur during the implosion once carbon was depleted
in the central zone.  Here, for the standard cases, time-dependent
mixing-length convection is allowed to proceed in zones that are not
supersonically collapsing, i.e., $v_{\rm col} < c_{\rm s}$, but the
convective speed used in the mixing length calculation is limited, in
all zones, to $v_{\rm conv} < 0.2\, c_{\rm s}$. Here $c_{\rm s}$ is the local sound
speed. Additional studies in which the zonal collapse speed limiter
was reduced to $v_{\rm col} < 0.3\, c_{\rm s}$ while retaining $v_{\rm conv} <
0.2\, c_{\rm s}$ showed little deviation from \Tab{pisn}. If those limiters were
further decreased though, to $v_{\rm col} < 0.1\, c_{\rm s}$ and $v_{\rm conv}
< 0.1\,c_{\rm s}$, \Mhi was reduced. More efficient mixing of oxygen
downwards during the implosion leads to greater burning and increases
\Mhi. The values in \Tab{pisn} are thus probably upper limits
for non-rotating models and the actual values could be 2\,\% smaller.
As noted previously, small changes in physics can have an large effect
near the cusp of a transition from full collapse to full
explosion. These changes thus had a much greater effect on \Mhi than \Mlo.

Rotation could also substantially decrease \Mhi by weakening
the pair instability, but that was not explored here. See
\Sect{rotation} for the effect of rotation on \Mlo.

\begin{deluxetable}{ccccc}
\tablecaption{Non-Rotating Helium Stars of Constant Mass}
\tablehead{ \colhead{${M_{\rm He}}$}         &
            \colhead{$X(^{12}$C$)$}          &
            \colhead{${f_{\rm Buch}}$}       &
            \colhead{${M_{\rm BH}}$}          &
            \colhead{${K\!E_{\rm exp}}$}
            \\
            \colhead{[\Msun]}              &
            \colhead{}                     &
            \colhead{}                     &
            \colhead{[\Msun]}              &
            \colhead{[10$^{51}$\,erg]}
            }\\
\startdata
50 & 0.0677 & 1.4 & 42.5 & 1.10 \\
51 & 0.0666 & 1.4 & 42.7 & 1.03 \\
52 & 0.0656 & 1.4 & 45.4 & 0.90 \\
53 & 0.0646 & 1,4 & 45.3 & 1.43 \\
54 & 0.0636 & 1.4 & {\bf 46.1} & 0.77 \\
56 & 0.0617 & 1.4 & 44.7 & 1.23 \\
58 & 0.0600 & 1.4 & 43.5 & 2.25 \\
60 & 0.0583 & 1.4 & 40.6 & 2.41 \\
62 & 0.0568 & 1.4 & 20.1 & 2.97 \\
64 & 0.0554 & 1.4 &  0   & 4.58 \\
   &        &     &       &      \\
50 & 0.0986 & 1.2 & 42.9  & 0.93  \\
51 & 0.0972 & 1.2 & 43.8  & 0.81  \\
52 & 0.0959 & 1.2 & 43.8  & 0.86  \\
53 & 0.0947 & 1.2 & 45.0  & 0.76  \\
54 & 0.0935 & 1.2 & 45.6  & 0.74  \\
56 & 0.0909 & 1.2 & 45.4  & 1.05  \\
58 & 0.0887 & 1.2 & {\bf 46.0} & 1.78  \\
60 & 0.0869 & 1.2 & 42.5  & 2.27  \\
62 & 0.0850 & 1.2 & 32.0  & 2.53  \\
64 & 9.0831 & 1.2 &  0    & 3.77   \\
   &        &     &       &        \\
50 & 0.140  & 1   & 44.9  & 0.41   \\
51 & 0.138  & 1   & 45.5  & 0.48   \\
52 & 0.137  & 1   & 46.1  & 0.47   \\
53 & 0.135  & 1   & 45.8  & 0.53   \\
54 & 0.134  & 1   & 46.5  & 0.46   \\
56 & 0.131  & 1   & {\bf 47.8}  & 0.67   \\
58 & 0.129  & 1   & 46.5  & 1.31   \\
60 & 0.126  & 1   & 46.2  & 1.75   \\
62 & 0.124  & 1   & 41.9  & 1.96   \\
64 & 0.122  & 1   & 24.1  & 2.70   \\
66 & 0.120  & 1   &  0    & 4.07   \\
   &       &     &       &        \\
50 & 0.187 & 0.8 & 49.0  &   0.05 \\
51 & 0.186 & 0.8 & 49.5  &   0.096 \\
52 & 0.184 & 0.8 & 49.9  &   0.14 \\
53 & 0.183 & 0.8 & 50.4  &   0.14 \\
54 & 0.181 & 0.8 & 51.0  &   0.15 \\
56 & 0.179 & 0.8 & 53.1  &   0.19 \\
58 & 0.176 & 0.8 & {\bf 56.5}  &   0.13 \\
60 & 0.173 & 0.8 & 53.4  &   0.90 \\
62 & 0.171 & 0.8 & 54.2  &   1.11 \\
64 & 0.169 & 0.8 & 47.0  &   1.74 \\
66 & 0.166 & 0.8 & 45.7  &   2.19 \\
68 & 0.164 & 0.8 & 31.0  &   2.71 \\
70 & 0.164 & 0.8 &  0    &   4.51 \\
   &       &     &       &        \\
54 & 0.190 & 1/1.35 & 50.9  & 0.21   \\
56 & 0.187 & 1/1.35 & 52.7  & 0.24   \\
58 & 0.184 & 1/1.35 & {\bf 56.6}  & 0.11   \\
60 & 0.181 & 1/1.35 & 52.9  & 0.67   \\
62 & 0.178 & 1/1.35 & 52.1  & 1.10   \\
64 & 0.176 & 1/1.35 & 52.3  & 2.16   \\
66 & 0.173 & 1/1.35 & 41.7  & 1.97   \\
68 & 0.171 & 1/1.35 & 13.0  & 3.37   \\
70 & 0.169 & 1/1.35 &   0    & 4.91   \\
   &       &        &        &        \\
56 & 0.261 & 0.75/1.35 &  55.9 & 0.001 \\
58 & 0.258 & 0.75/1.35 &  57.8 & 0.007 \\
60 & 0.254 & 0.75/1.35 &  59.6 & 0.020  \\
62 & 0.252 & 0.75/1.35 &  60.7 & 0.10   \\
64 & 0.249 & 0.75/1.35 & 63.1  & 0.042 \\
66 & 0.246 & 0.75/1.35 & 63.5  & 0.17   \\
68 & 0.244 & 0.75/1.35 & {\bf 64.1} & 0.32   \\
70 & 0.241 & 0.75/1.35 & 51.4 & 1.39   \\
72 & 0.239 & 0.75/1.35 & 25.5 & 2.85   \\
74 & 0.236 & 0.75/1.35 & 11.6 & 3.95   \\
76 & 0.234 & 0.75/1.35 &   0   & 4.53
\enddata
\tablecomments{The S-factor used for the
  $^{12}$C($\alpha,\gamma)^{16}$O reaction rate in these calculations
  was 146\,keV\,b multiplied the indicated multiplication factor,
  $f_{\rm Buch}$. In some runs the $3\alpha$ rate was also multiplied
  by 1.35. $X$($^{12}$C) is the central carbon mass fraction just
 before central carbon burning ignites.}  \lTab{helium}
\end{deluxetable}

\begin{deluxetable}{ccccc}
\tablecaption{Upper Bound on Pair-Instability Supernovae}
\tablehead{ \colhead{${M_{\rm He}}$}          &
            \colhead{${f_{\rm Buch}}$}         &
            \colhead{$X(^{12}$C$)$}          &
            \colhead{${M_{\rm BH}}$}          &
            \colhead{${K\!E_{\rm exp}}$}
            \\
            \colhead{[\Msun]}              &
            \colhead{}                     &
            \colhead{}                     &
            \colhead{[\Msun]}              &
            \colhead{[10$^{51}$\,erg]}
            }\\
\startdata
135 & 1.4 & 0.0311 &  0   & 99.4  \\
136 & 1.4 & 0.0309 & 136  &   0   \\
    &     &        &      &       \\
138 & 1.2 & 0.0487 &  0   & 103   \\
139 & 1.2 & 0.0485 & 139  &   0   \\
    &     &        &      &       \\
143 & 1.0 & 0.0758 &  0   & 109   \\
144 & 1.0 & 0.0755 & 144  &   0   \\
    &     &        &      &       \\
150 & 0.8 & 0.116  &  0   & 116   \\
151 & 0.8 & 0.116  & 151  &   0   \\
    &     &        &      &       \\
151 &  1.0/1.35  & 0.115  &  0  & 120 \\
152 &  1.0/1.35  & 0.114  & 152 &  0  \\
    &            &        &     &     \\
160 &  0.75/1.35 & 0.176  &  0  & 131 \\
161 &  0.75/1.35 & 0.176  & 161 &  0  \\
\enddata
\lTab{pisn}
\end{deluxetable}

\subsection{Helium Stars With Mass Loss}
\lSect{mdot}

Stellar winds reduce the mass of the star and alter the thickness of
the helium burning shell relative to the carbon-oxygen core.  Being
farther out in the star, the influence of the residual helium is
reduced relative to carbon, but helium burns at a lower temperature
and generates more energy, and so can still have a minor effect. Stars
with the large masses considered here have convective helium burning
cores that extend through more 90\,\% of their mass when helium is half
burned, so most of the post-helium burning star is carbon and oxygen.
Some helium always remains in the outer layers of the star though, no
matter what the mass loss rate \citep[see, e.g., Table~4
  of][]{Woo19}. The convective core recedes as mass is lost from the
surface leaving a gradient of helium outside.  Typically at carbon
ignition helium still has an appreciable abundance in the outer 15\,\%
of the star. Burning at the base of this shell during implosion drives
convection that mixes additional helium down, increasing the strength
of the shell burning.

To illustrate the effect, we consider here a single set of reaction
rates, $f_{\rm Buch} = 1.2$ and $f_{3\alpha}$ = 1.0, and explore
some initial mass and mass loss combinations that produce presupernova
stars in the same mass range as \Tab{helium}.  The mass loss rate is
taken from \citet{Yoo17} for 10\,\% solar metallicity with $f_{\rm WR} =
1$. \citet{Vin17} and \citet{San19} predict a smaller mass loss rate
for such stars.  \Tab{wind} shows some results. For a typical case, a
helium core with an initial mass of 76\,\Msun\ ends up with a final
mass of 55.9\,\Msun, mostly composed of carbon and oxygen, with a
central carbon fraction of 0.0862. This is very similar to the 56\,\Msun\ helium star which, when evolved at constant mass in
\Tab{helium}, had a carbon mass fraction of 0.0909. The small
difference is because the shrinking helium convective core mixes in
less helium to the center in the mass-losing star towards the end of
helium burning. A more important difference though is the amount of
helium in the outer regions of the star. The model with mass loss
had 0.94\,\Msun\ of $^{4}$He prior to pair pulsations, while the
constant mass model had 1.94\,\Msun. Both helium shells had their bases
at about 47\,\Msun\ though, i.e., the outer 9\,\Msun\ of both
presupernova stars contained appreciable helium.

\Tab{wind} shows that the effect of this residual helium, and mass loss
on the remnant mass is generally small for the reaction
rates considered. Comparing the results including mass loss to the
corresponding stars with $f_{\rm Buch}=1.2$ and $f_{3 \alpha} = 1$
in \Tab{helium}, \Mlo is reduced by less than 1\,\Msun. This
is comparable to variations introduced by zoning, network, and
time step criteria.  The small effect of the helium shell also
suggests that the essential conclusions of \Tab{helium} might be
obtained ignoring the helium altogether and just modeling stars of
just carbon and oxygen.

\begin{deluxetable}{ccccc}
\tablecaption{Effect of Mass Loss}
\tablehead{ \colhead{${M_{\rm init}}$}          &
            \colhead{${M_{\rm fin}}$}         &
            \colhead{$X(^{12}$C$)$}          &
            \colhead{${M_{\rm BH}}$}          &
            \colhead{${K\!E_{\rm exp}}$}
            \\
            \colhead{[\Msun]}              &
            \colhead{[\Msun]}              &
            \colhead{}                     &
            \colhead{[\Msun]}              &
            \colhead{[10$^{51}$\,erg]}
            }\\
\startdata
68 & 50.2 & 0.0931 & 42.7 & 1.27  \\
70 & 51.6 & 0.0913 & 45.2 & 0.66  \\
72 & 53.0 & 0.0895 & 44.8 & 0.62  \\
74 & 54.4 & 0.0878 & 45.7 & 0.58  \\
76 & 55.9 & 0.0862 & 45.5 & 0.86  \\
78 & 57.2 & 0.0846 & 45.0 & 1.41  \\
80 & 58.7 & 0.0831 & 41.0 & 1.47
\enddata
\lTab{wind}
\end{deluxetable}

\subsection{Pure Carbon-Oxygen Stars}
\lSect{carbonstar}

Many factors affect the carbon abundance in the core of a massive star
following helium burning: the reaction rates for $3 \alpha$ and
$^{12}$C($\alpha,\gamma)^{16}$O; the treatment of semi-convection and
convective overshoot mixing; rotationally induced mixing; radiative
mass loss (\Sect{mdot}); and whether the star is in a mass exchanging
binary \citep{Woo19}. Fortunately, the structure of a very massive
star during its final stages of helium burning is quite
simple. Because the luminosity is nearly Eddington and the structure
near that of an $n=3$ polytrope, helium stars of the mass considered
are almost fully convective throughout most of their helium-burning
evolution. Their compositions and entropies, except for a small mass
near the surface, are nearly constant. The helium exhausted
material includes most of the matter that participates in burning
during a PPISN or PISN. The remainder, near the surface, while
composed of a mixture of helium and carbon, has an electron mole
number near $Y_{\rm e} = 0.50$. The star's structure, which is most
sensitive to the electron density, is thus not very much affected by
the ratio of helium to carbon.

These characteristics and a desire to better understand the star's
composition affects \Mlo motivate an exploration of stars
consisting initially of just carbon and oxygen in a ratio that does
not vary with location in the star. The composition reflects what
exists at the end of helium burning and condenses many of the
uncertainties mentioned above into a single parameter, the carbon mass
fraction. Once a grid of explosions has been calculated for a suitable
range of carbon mass fractions, the future properties of the supernova
can determined for any model by comparing its actual carbon mass
fraction at helium depletion to the grid. Since the carbon mass
fraction is bounded, so too is the maximum mass black hole that can be
produced following the pulsational-pair instability.

Unfortunately, such a simple approach suffers from two
complications. First, as discussed in \Sect{mdot}, helium shell
burning can weaken the pair instability. Whether the outer layers of
the star are carbon or helium makes little difference to their
structure, but the burning of helium during the implosion can weaken
the first pulse. As a consequence, the remnant masses coming out of
pure carbon-oxygen cores could underestimate the actual value of \Mlo. The other complication is that helium burning in such massive
stars produces not just $^{12}$C and $^{16}$O, but appreciable
$^{20}$Ne and even some $^{24}$Mg as well. The abundances of these
heavier nuclei can become comparable to the carbon abundance itself
for those cases where the $^{12}$C($\alpha,\gamma)^{16}$O reaction
rate is large and the $3 \alpha$ rate small. For the converse cases,
perhaps of greater interest here, where the carbon abundance after
helium burning is large, the carbon shell dominates and both the
burning of helium and the abundance of neon have little effect. The
effects of neon are also mitigated by the fact that neon burning never
becomes an exoergic phase in such massive stars. Neutrino losses
dominate and the products of neon burning are oxygen and magnesium, so
half the $^{20}$Ne ends up in $^{16}$O anyway. The bulk energy yields
of the neon-rich and neon-poor compositions also differ very
little. Burning the ``0.05/Ne'' composition in \Tab{cstar} to
$^{28}$Si gives $4.63 \times 10^{17}$\,erg\,g$^{-1}$ whereas burning the
``0.05'' composition to $^{28}$Si gives $4.69 \times 10^{17}$\,erg\,g$^{-1}$.  Nevertheless, having a substantial abundance of neon can
sometimes affect the strength of carbon shell burning and it will be
necessary to demonstrate the sensitivity of the answer to the neglect
of neon and magnesium in the initial composition and of helium in the
outer layers.

\Tab{cstar} gives the results. Carbon-oxygen stars of constant mass
with the initial carbon mass fractions indicated were evolved until the
star either completely disperses as a PISN or its iron core
collapses. Except for the three columns noted, the remainder of the
initial star was all $^{16}$O. The range of carbon mass fractions span
what we consider to be reasonable values for current reaction rates
and convective treatments. No helium star in \Sect{helium} gave a
central carbon mass fraction less than 0.05 or greater than 0.25. In
two of the exceptional series of models, those marked ``0.05/Ne'' and
``0.1/Ne'', more realistic neon-rich compositions were considered.
Helium star models in \Tab{helium} show that when the carbon mass
fraction after helium burning is 0.05, the mass fractions of $^{20}$Ne
and $^{24}$Mg are 0.07 and 0.01 respectively. When carbon is 0.1,
$^{20}$Ne and $^{24}$Mg are 0.05 and 0.005. Surveys using these
``neon-rich'' compositions are also shown for comparison in
\Tab{cstar}.

The effects of a helium shell were also considered in a series of
models labeled ``0.05/He''.  In the inner 85\,\% of their mass, these stars had
compositions of 5\,\% carbon and 95\,\% oxygen by mass. In their outer
15\,\% though the composition was 5\,\% carbon, 90\,\% oxygen and 5\,\%
helium. This typically amounted to about 0.4\,\Msun\ of helium, small
even for mass losing stars (\Sect{mdot}). As expected the presence of
helium generally led to larger black hole masses. The effects of both
neon and helium in the ``0.05'' case was an increase of about 2\,\Msun
\ (5\,\%) each in \Mlo. The ``0.05'' case should give an upper
bound on these effects because the carbon plays an increasingly
dominant role in the others.


Comparing \Tab{cstar} for carbon-oxygen stars with \Tab{helium} for
helium stars that have similar carbon mass fractions at helium
depletion, one sees that discrepancy is never great, but is least when
the carbon-mass fraction is large. Not only is the neon abundance
smaller in such cases, but the effects of carbon shell burning
dominate those of helium shell burning. Conversely, when the carbon
abundance is small, neon and helium have a bigger effect. For For
$X(^{12}$C) = 0.05, \Mlo is 41\,\Msun\ for carbon-oxygen
stars, but 46\,\Msun\ for helium stars with $f_{\rm Buch} =
1.4$. About half the difference is due to the large neon mass fraction
and the other half is due to the helium shell (columns ``0.05/He'' and
``0.05/Ne'').  For $X(^{12}$C) = 0.125, on the other hand, \Mlo is 47\,\Msun\ for carbon-oxygen stars and 48\,\Msun\ for
helium stars with $f_{\rm Buch} = 1$. For $X(^{12}$C) = 0.25, \Mlo is 64\,\Msun\ for carbon-oxygen stars and 64\,\Msun\ for
helium stars with $f_{\rm Buch} = 0.75$ and A $3 \alpha$ rate 1.35
times standard. Variations of 1 - 2\,\Msun\ are expected in any survey
due to zoning, time step, network, number of pulses, etc. Thus unless
the carbon abundance is less than about 0.1, the simpler carbon-oxygen
cores can be substituted for helium stars in our search for \Mlo.

As with the helium stars, the maximum black hole mass in \Tab{cstar}
is not very sensitive to the exact value of the initial fraction of
carbon so long as that fraction is small. For carbon fractions from
0.05 to 0.15, \Mlo varies only between 41 and 48\,\Msun\ (and
41\,\Msun\ is an underestimate as we have just discussed). For larger
values though, as would occur for small values of the
$^{12}$C($\alpha,\gamma)^{16}$O reaction cross section or a large
value for triple-alpha, \Mlo increases dramatically.

To summarize, for a $S$-factor greater than 110\,keV\,b \citep{Deb17} and
a triple-alpha rate no greater than 1.35 times the traditional value,
the carbon mass fraction in our models is no greater than 0.25 at
helium depletion. This robustly implies a most massive black hole
below the pair instability gap of $\Mlo = 64$\,\Msun\ with an
expected error of about 2\,\Msun. Pending further experiments, we
regard this as the upper limit to the black hole mass one can obtain
in a close mass exchanging binary by adjusting only the nuclear
reaction rates. Additional corrections for mass loss are small
(\Tab{wind} vs.\ \Tab{helium}).

\begin{deluxetable*}{ccccccccccccc}
\tablecaption{Remnant Mass in Carbon-Oxygen Stars as a Function of $^{12}$C Mass Fraction}
\tablehead{ \colhead{${M_{\rm CO}}$}         &
            \colhead{0.05}        &
            \colhead{0.05/He}     &
            \colhead{0.05/Ne}     &
            \colhead{0.075}       &
            \colhead{0.10}        &
            \colhead{0.10/Ne}     &
            \colhead{0.125}       &
            \colhead{0.15}        &
            \colhead{0.175}       &
            \colhead{0.20}        &
            \colhead{0.225}       &
            \colhead{0.25}
            \\
            \colhead{[\Msun]}      &
            \colhead{}      &
            \colhead{}      &
            \colhead{}      &
            \colhead{}      &
            \colhead{}      &
            \colhead{}      &
            \colhead{}      &
            \colhead{}      &
            \colhead{}      &
            \colhead{}      &
            \colhead{}      &
            \colhead{}
            }\\
\startdata
44  &  37.1  &  37.3  &  37.4  &    -   &   -    &   -    &     -    &   -    &   -   &  -     &  -   &  -  \\
46  &  38.7  &  39.4  &  39.2  &  39.5  &   -    &   -    &     -    &   -    &   -   &  -     &  -   &  -  \\
48  &  40.1  &  41.7  &  41.0  &  41.9  &  41.5  &  41.6  &     -    &   -    &   -   &  -     &  -   &  -  \\
50  &  40.5  &  42.4  &  43.2  &  43.5  &  43.8  &  43.3  &   43.5   &   -    &   -   &  -     &  -   &  -  \\
52  &  40.6  &  43.1  &  41.7  &  43.3  &  45.5  &  44.8  &   44.9   &  47.4  &   -   &  -     &  -   &  -  \\
54  &  37.2  &  42.0  &  40.7  &  35.8  &  44.0  &  44.4  &   46.0   &  47.9  &  50.2 &  51.8  &  -   &  -  \\
56  &  24.7  &  38.2  &  35.1  &  35.4  &  42.3  &  45.1  &   47.2   &  48.2  &  50.4 &  53.0  & 55.1 &  -  \\
58  &   0    &   0    &    0   &  16.7  &  38.6  &  26.0  &   43.1   &  47.7  &  51.3 &  53.1  & 55.0 & 57.6 \\
60  &   -    &    -   &    -   &   0    &  19.5  &  22.5  &   37.5   &  46.9  &  49.9 &  56.7  & 58.9 & 58.7 \\
62  &   -    &    -   &    -   &    -   &   0.0  &   0.0  &   19.2   &  33.4  &  48.8 &  55.2  & 57.3 & 59.2 \\
64  &   -    &    -   &    -   &    -   &    -   &   -    &     0    &  23.6  &  44.2 &  51.6  & 61.5 & 62.3 \\
66  &   -    &    -   &    -   &    -   &    -   &   -    &     -    &    0   &  31.9 &  43.9  & 57.7 & 62.4 \\
68  &   -    &    -   &    -   &    -   &    -   &   -    &     -    &   -    &   0   &  33.1  & 52.3 & 63.7 \\
70  &   -    &    -   &    -   &    -   &    -   &   -    &     -    &   -    &   -   &   0    & 38.3 & 56.0 \\
72  &   -    &    -   &    -   &    -   &    -   &   -    &     -    &   -    &   -   &   -    & 18.9 & 39.9 \\
74  &   -    &    -   &    -   &    -   &    -   &   -    &     -    &   -    &   -   &   -    &  0.  & 20.2 \\
76  &   -    &    -   &    -   &    -   &    -   &   -    &     -    &   -    &   -   &   -    &  -   &  0   \\
78  &   -    &    -   &    -   &    -   &    -   &   -    &     -    &   -    &   -   &   -    &  -   &  -   \\
    &        &        &        &        &        &        &          &        &       &        &      &      \\
\Mlo &   41   &   43   &   43   &   44   &   46   &   45   &    47    &   48   &   51  &    57  &  61  &  64
\enddata

\tablecomments{Remnant masses are given for carbon-oxygen stars of
  initially constant composition with the mass fraction of carbon
  indicated. Models 0.05/Ne had an initial composition of 5\,\%
  $^{12}$C, 87\,\% $^{16}$O, 7\,\% $^{20}$Ne and 1\,\% $^{24}$Mg. Models
  0.1/Ne had an initial composition of 10\,\% $^{12}$C, 84.5\,\% $^{16}$O,
  5\,\% $^{20}$Ne and 0.5\,\% $^{24}$Mg.  Models 0.05/He had a initial
  mass fraction of helium of 0.05 in the outer 15\,\% of their mass. The
  $^{16}$O mass fraction was reduced there by 0.05 to compensate.  For
  all other models the mass that was not $^{12}$C was $^{16}$O. The
  maximum value for each carbon mass fraction is \Mlo. }
\lTab{cstar}
\end{deluxetable*}





\section{Single Stars}
\lSect{single}

More massive black holes can result from presupernova stars that
retain their hydrogen envelope. It is assumed that the envelope
participates, along with the helium core in the collapse to a black
hole. Very massive single stars could die with most of their envelope
intact if their metallicity is small enough that their mass loss is
negligible. The envelope might also be retained in a wide detached
binary systems. Both assumptions have their problems. Even if
radiative driven winds retain their proposed scalings for the very
high masses and low metallicities where they have not been tested
experimentally, stars near the Eddington limit may find other ways of
losing mass. The mass loss rates for luminous blue variables and stars
as massive as Eta Carina are very uncertain, and might not diminish
greatly just because the radiative opacity is less \citep{Smi06}.
Binary systems so wide as to contain supergiants may also have
difficulty merging by gravitational radiation in a Hubble time.

\begin{deluxetable*}{cccccccc}
\tablecaption{Single Star Models}
\tablehead{ \colhead{Model}                &
            \colhead{${f_{\alpha,\gamma}/f_{3 \alpha}}$}         &
            \colhead{${M_{\rm preSN}}$}       &
            \colhead{${M_{\rm He}}$}         &
            \colhead{${X(^{12}C)}$}         &
            \colhead{${M_{\rm SCZ}}$}        &
            \colhead{${M_{\rm BH}}$}         &
            \colhead{${K\!E_{\rm exp}}$}
            \\
            \colhead{       }              &
            \colhead{       }              &
            \colhead{[\Msun]}              &
            \colhead{[\Msun]}              &
            \colhead{}                     &
            \colhead{[\Msun]}              &
            \colhead{[\Msun]}              &
            \colhead{[10$^{51}$\,erg]}
            }\\
\startdata
T90Ca  & 0.8/1.0  & 80.92  & 41.13 & 0.202 & 55.2 &  80.9    & $<0.0001$ \\
T100Ca & 0.8/1.0  & 88.50  & 46.53 & 0.198 & 65.3 &  49.9    & 0.151     \\
T100Cb & 0.8/1.35 & 88.79  & 46.53 & 0.262 & 65.3 &  88.7    & $<0.0001$ \\
T110Cb & 0.8/1.35 & 97.13  & 50.57 & 0.254 & 71.5 & $\sim73$ & 0.021    \\
T120Cb & 0.8/1.35 & 104.76 & 55.17 & 0.249 & 77.3 & $\sim67$ & 0.077    \\
T130Cb & 0.8/1.35 & 112.27 & 60.37 & 0.239 & 77.6 &  63.5    & 0.173    \\
T140Cb & 0.8/1.35 & 119.31 & 66.59 & 0.232 & 78.8 &  65.0    & 0.65
\enddata
\tablecomments{The model name gives the zero age main sequence mass of
  the star. $M_{\rm SCZ}$ is the mass of the presupernova star that is
  inside the base of the surface convection zone. The external binding
  energy there is close to 10$^{48}$\,erg. $M_{\rm He}$ is the helium
  core mass and $M_{\rm preSN}$, the total presupernova mass. All
  quantities except the remnant mass, $M_{\rm BH}$, and final kinetic
  energy, $K\!E_{\rm exp}$, are evaluated at carbon ignition.  }
\lTab{single}
\end{deluxetable*}


Still, it is interesting that the larger carbon abundances allowed by
altered reaction rates also imply that single stars with larger total
mass could collapse. \citet{Bel10} estimated a maximum black hole mass
of 80\,\Msun\ for a metallicity 1\,\% of solar, but did not include the
effect of the pair instability.  \citet{Woo17} found that a black hole
of 65\,\Msun\ resulted from the evolution of a 70\,\Msun\ star with a
a very low mass loss rate (Model T70C). That calculation was for a
relatively large rate for $^{12}$C($\alpha,\gamma)^{16}$O ($f_{\rm
  Buch} = 1.2$) and a standard value for 3$\alpha$.  Whereas not
attempting a full survey, we consider here the evolution of six
metal-deficient massive stars, Models T90C, T100C, T110C, T120C,
T130C, and T140C \citep[see Table 2][]{Woo17}, using variations on the
reaction rates chosen to produce large central carbon abundances at
helium depletion. These models did not rotate and had zero age main
sequence masses of 90 to 140\,\Msun. Their metallicity was 10\% solar,
but their mass loss rate as main sequence stars and red supergiants
was further reduced by an additional factor of 8 compared to the
standard value for that metallicity. This would require a still lower
value of metallicity or mass loss rates much less than standard values
for 10\,\% solar metallicity.

Final presupernova masses, helium core masses, and remnant masses are
given in \Tab{single} along with the variations in reaction rates and
resulting carbon mass fraction at helium depletion.  The behavior of
these full star explosions generally follows from \Tab{helium} and
\Tab{cstar}, but due to the non-negligible boundary pressure of the
hydrogen shell, the equivalent helium core mass in \Tab{single} is
\ smaller than for the bare helium star models. Given their large
carbon mass fractions, the pair instability in Models T90Ca and T100Cb
is mild and launches weak shocks having initial speeds of only 100's
of km\,s$^{-1}$. These shocks appear late in the evolution after stable
central oxygen burning and most of silicon core burning have already
transpired. Typical explosion energies were 10$^{47}$\,erg or
less. This only suffices to eject a small amount of material near the
surface, even in a red supergiant, so the remnant mass very nearly
equals the presupernova mass.

Other models experienced stronger pulsations, though still late in
their evolution. The presupernova models were all characterized by
large radii, $\gtaprx 10^{14}$ cm, but had surface convection zones
that had not dredged up the entire envelope. The net binding energy of
the dredged up material for all six stars was, to within about a
factor of two, $\sim10^{48}$\,erg, while the binding of all matter
external to the helium core, including the part that was not
convective, was greater that $\sim10^{50}$\,erg. Thus when explosions
developed with energy of order 10$^{49}$\,erg, typical mass separations
lay beneath, but not far beneath the base of the convective part of the
envelope (Models 110Cb and 120Cb). These mass separations were
difficult to determine because shock speeds were so slow that fallback
continued for decades following the initial collapse. The location of
the base of the convective zone is also model dependent since it is
sensitive to the convection model, the opacity and metallicity, and
mass loss. The depth of this zone was also slightly different at helium
depletion, carbon ignition (where the evaluations in \Tab{single} were
made), and carbon depletion. Had the model been a luminous blue
variable or blue supergiant, the binding of the envelope might have
been greater and so would the remnant mass, but then mass loss for the
luminous blue variable could have been large. For larger explosion
energies (Models 100Ca, 130Cb and 140Cb) the mass cut is more
precisely determined. For all models except T140Cb, the pulses
occurred so late that the iron core collapsed while the outgoing shock
from the pulses was still inside the star.

For the small values of $^{12}$C($\alpha,\gamma)^{16}$O rate and large
$3 \alpha$, black holes of 85\,\Msun\ and more are possible. These are
cases where the advanced stages of evolution are essentially stable
and pulsational mass loss is negligible (Models T90Ca and
100Cb). Since the helium core properties are largely independent of
the mass loss rate, unless the whole envelope is lost, somewhat larger
black hole masses could be achieved given ones tolerance for reducing
the mass loss rate, but probably not much beyond 100\,\Msun.  It is
interesting that black holes of such high mass can form, even in
isolation. If they were to some day be captured in another system, one
would not need to resort to multiple mergers to build up a black hole
of 85\,\Msun.

As noted before, there might be problems with making a LIGO source
directly this way. Given the large radii of the stars, bringing the resulting
black hole from the first explosion close enough to the secondary in
order to merge in a Hubble time requires additional assumptions. One
possibility is that the primary (which makes the more massive black
hole) was a blue supergiant or luminous blue variable and the
secondary a red supergiant when they died. The larger radius for the
secondary might provoke mass exchange that was avoided for the
primary.  This might happen e.g., if the secondary was a low
metallicity star that made primary nitrogen when its helium convective
core encroached on the hydrogen envelope while the primary did
not. Such evolution is often seen for stars of this large mass and low
metallicity \citep[e.g.,][]{Heg10}.

Another possibility is kicks during the explosion \citep{Tut17}. The
difficulty here is that if no mass is ejected in a jet, the kick must
be a consequence of asymmetric neutrino emission, but neutrino losses
are a small fraction of the stellar mass. Once a black hole forms,
without rapid rotation, the infalling material does not radiate
neutrinos effectively. If neutrino emission is restricted to just the
phase of proto-neutron star evolution, then the losses are probably no
more than 0.5\,\Msun\,c$^2$. Given a 5\,\% asymmetry, which seems liberal,
the kick for an 80\,\Msun\ black hole would then be $(0.5)(0.05)/80\,c
\sim 100$\,km\,s$^{-1}$.  These neutrinos would need to be
coincidentally radiated largely in the direction of orbital
motion. This could cancel the orbital momentum for a wide enough
separation, but the balance would need to be nearly precise to result
in a plunging highly eccentric orbit that might later be circularized
if it passed close enough to the secondary. This seems feasible but
unlikely.

\section{Rotation}
\lSect{rotation}

\subsection{Formalism and Physics}
\lSect{rotphys}

Rotation weakens the pair instability \citep{Fow64}, but the effect on
the dynamics of the explosion is not great unless the star maintains a
large degree of differential rotation.  \citet{Mar20} give the
modification to the structural adiabatic index, $\Gamma_{1}$, required for
instability,
\begin{equation}
  \Gamma_{1} \ < \ \frac43 \, - \, \frac29\,\omega_{\rm k}^2 +
  \mathcal{O}(\omega_{\rm k}^4),
\lEq{instability}
\end{equation}
where $\omega_{\rm k} = \Omega/\sqrt{G m /r_{\rm e}^3}$ is the ratio of the
actual angular rotation rate, $\Omega$, to the Keplerian rotation rate
for a given mass shell; $r_{\rm e}$ is its equatorial radius, here just
taken to be the radius; and $m$ is the mass inside that radius. Since
a real star has neither constant $\Gamma_1$ or $\omega_{\rm k}$, this
criterion may or may not be satisfied in different regions of the star
at different times. Whether there is a global instability depends upon
the integrated average \citep{Mar20}.

The strength of the pair instability also depends sensitively upon the
temperature and is weaker when the star first begins to contract and
gain momentum, and greater at the end when oxygen burns
explosively. The accumulated momentum during the collapse plays an
important dynamical role.  Typical maximum decrements in $\Gamma_1$
from the pair instability for the masses being studied, once the
central temperature exceeds $2 \times 10^9$\,K, are a few percent
locally (relative to a fiducial 1.33), and $\sim$1\,\% globally
(\Fig{rotcore}). This implies that differential rotation in the deep
interior with $\omega_{\rm k} \gtaprx$ 10\,\% will be required to
alleviate the instability.  Smaller amounts of rotation can still
affect the outcome though by leveraging the evolution early on when
the instability is weak. \citet{Mar20} found a typical increase in
black hole mass of $\sim$4\,\% when magnetic torques \citep{Spr02}
were included in the models, and $\sim$15\,\% when they were
not. Since ignoring magnetic torques would give a very rapid rotation
rate for pulsars and possibly overproduce gamma-ray bursts
\citep{Woo06}, the inclusion of magnetic torques is probably more
realistic.

Unfortunately, rapid rotation affects the evolution of massive stars
in many ways, and disentangling them complicates any conclusion
regarding the black hole birth function. Rotation leads to mixing that
results, at a minimum, in larger helium cores for a given main
sequence mass. At a maximum, the star may become fully mixed on the
main sequence and experience chemically-homogeneous evolution
\citep{Mae87}.  Rotationally-induced mixing can transport additional
helium into the convective helium burning core at late times, thus
decreasing the carbon yield and lowering the remnant mass
(\Tab{cstar}). Rotation reduces the central density making the star
evolve like one of lower mass \citep{Mar20}. Rapid rotation can lead
to mass shedding at the equator, especially in stars that are already
near the Eddington limit - as these are. Rotation can mix helium or
carbon down during the implosion so that shell burning has a greater
effect during the first pulse. Rotation can alter the way the
collapsing iron core behaves and allow explosions to develop where
the star would have collapsed if neutrino transport were the only
mechanism. If the rotation of the remnant is too rapid, not all of the
star can immediately collapse to a black hole without making a
disk. The formation of a ``collapsar'' may produce a jet that explodes
the remaining star or, at the least, inhibits its full collapse.

The amount and distribution of angular momentum also depends
critically upon uncertain mechanisms for its transport and the rates
for mass loss.  Transport concentrates angular momentum in the outer
layers of the star where mass loss removes it. To neglect mass loss is
to overestimate the role of rotation in the evolution, but to include
it adds another uncertain parameter to the outcome, a parameter that
will itself vary in an uncertain way with metallicity.

For all these reasons, the results in this section are less precise
than those in the previous ones. A one-dimensional code is used to
calculate the effects of a two-dimensional phenomenon -
rotation. Angle averaging and a semi-empirical formulation is
necessary.  Mass loss by winds is neglected, though see
\Sect{prog}. What is most important is the angular momentum
distribution the star has at carbon ignition since radiation-driven
mass loss can presumably can be neglected after that. Though we follow
all stars through helium burning, they approach carbon ignition with
essentially the same angular momentum distribution they started with -
nearly rigid rotation characterized by a single parameter, $J_{\rm
  init}$, the initial total angular momentum of the star.

Rotational mixing and angular momentum transport by the usual
dynamical processes \citep{Heg00} were included in all stages of the
evolution, including the inter-pulse periods. Time-dependent
convection is also left on during inter-pulse periods in that portion
of the star that remains in hydrostatic equilibrium, i.e., is not
moving supersonically.  Except where specified, angular momentum
transport by magnetic torques is included \citep{Spr02,Heg05}.

Unlike our past studies of pair instability, an approximate
centrifugal support term has been added to the momentum equation
as \begin{equation}f\frac94\,\frac{j^2}{r^{3}}\approx
  f\,\frac{\Omega^2}r\end{equation} where $j$ is the average specific
  angular momentum of the shell, $r$ is the radius of the shell, and
  $\Omega$ is the angular velocity of the shell.  Shells are assumed
  to be spherical and rigidly rotating (solid body rotation for each
  shells).  Different shells may rotate at different angular velocity.
  The factor $\nicefrac94$ is because, for a spherical shell in solid
  rotation, the average angular momentum at the equator is
  $\nicefrac32$ times the average.  The factor $f$ accounts for
  averaging the centrifugal force over the surface, and should be
  $\nicefrac23$.  Our approximation does not take into account any
  deformation of the shell or that the centrifugal forces would be
  larger at the equator than at the pole.

The helium stars studied here are assumed to initially rotate rigidly and are
characterized by a total angular momentum, $J_{\rm init}$ = 0.15, 0.3,
0.6, and $1.2\times 10^{53}$ \,erg\,s. Several choices of reaction rates
are explored. Except for rotation, the physics and computational
procedure here are the same as for non-rotating helium stars in
\Sect{helium}.

Rotational mass shedding is included and is important for many models
with $J_{\rm init} \gtaprx\; 6\times 10^{52}$\,erg\,s. A surface zone is
assumed to be lost when its angular velocity exceeds 50\,\% of Keplerian
(i.e., $\omega_{\rm k} > 0.5$). This value, corresponding to a ratio of
centrifugal force to gravity of 25\,\%, is an upper bound since the
luminosity of the stars considered here is typically $\sim90\,\%$
Eddington. In any case, a larger value would make the approximation of
spherical stars highly questionable. Values of $\omega_{\rm k}$ this large
only exist in layers near the surface. The central density rises from
around 1,000\,g\,cm$^{-3}$ prior to carbon ignition ($T_{\rm 9,c} = 0.5$) to
10$^{7}$\,g\,cm$^{-3}$ during silicon burning. As the star attempts to
maintain rigid rotation, the surface of the star is spun up, resulting
in mass loss. This loss continues all the way through the evolution.

In addition to their angular momentum and mass, the models in
\Tab{rotate} are characterized by several choices of nuclear reaction
rates chosen to span most of the range for the non-rotating stars in
\Sect{helium} and \Tab{helium}. Three choices typify a central carbon
mass fraction at carbon ignition, X($^{12}$C), that is either
``small'', $f_{\rm Buch} = 1.2$, $f_{3 \alpha} = 1$; ``medium'',
$f_{\rm Buch} = 1.0$, $f_{3 \alpha} = 1.35$; or ``large'', $f_{\rm
  Buch} = 0.75$, $f_{3 \alpha} = 1.35$. Magnetic torques are included
all the time in all models except 64b, 66b, 68b, and 80b.

The rotating models in \Tab{rotate} will be denoted by their mass and
the choice of reaction rates and angular momentum used in their
calculation. Model 60$_{1.2,12}$ thus refers to the 60\,\Msun\ model
that used $f_{\rm Buch} = 1.2$, $f_{3 \alpha} = 1$, and an initial
angular momentum $J_{\rm init} = 12 \times 10^{52}$\,erg\,s.

\subsection{Results for Rotating Helium Stars}
\lSect{rotresult}

Model stars (\Tab{rotate}) were generated by allowing an extended
configuration of pure helium with an initial central density of
0.1\,g\,cm$^{-3}$ to relax to thermal and hydrostatic equilibrium with
a central density of about 200\,g\,cm$^{-3}$ and central temperature
$\sim2.2 \times 10^8$\,K where helium burning ignited. The initial
angular momentum, $J_{\rm init}$, characterizes that state. Helium
burning lasts about 300,000 years in all models and produces the
central carbon mass fractions given in the table. Typical radii at
central helium depletion (helium mass fraction is 1\%) are near $1.5
\times 10^{11}$\,cm and the luminosities are $\sim1.05 \times
10^{40}$\,erg\,s$^{-1}$ \ (M / 60\,\Msun).

After helium depletion, carbon burns away at the centers of all models
without the nuclear energy generation ever exceeding neutrino losses
and generating convection. Due to efficient magnetic torques, the star
rotates nearly rigidly up until carbon ignition except for a small
mass at the surface. The angular velocity for most of the interior is
$\Omega \approx 5 \times 10^{-4} (60\,\Msun/M)^2 (J_{\rm init,52}/3)$
radians s$^{-1}$ when the central temperature is $5 \times 10^8$\,K
and density, 2,500\,g\,cm$^{-3}$. The ratio of angular velocity to its
Keplerian value, $\omega_{\rm k}$, varies with radius, but for a point
midway in mass in the star is typically $\omega_{\rm k} \approx 0.029
(60\,\Msun/M)^2 (J_{52,0}/3)$ at that time (\Fig{rotcore}). These
values are reduced in cases where there has been appreciable rotational mass
shedding.

\begin{deluxetable*}{cccccccccc}
\tablecaption{Helium Stars with Constant Mass and Rotation}
\tablehead{ \colhead{{$M_{\rm He}$}}      &
            \colhead{${f_{\rm Buch}}$}     &
            \colhead{{$J_{\rm init}$}}     &
            \colhead{{$X(^{12}$C)}}      &
            \colhead{{$M_{\rm shed}$}}     &
            \colhead{{$M_{\rm pulse 1}$}}   &
            \colhead{{$K\!E_{\rm pulse 1}$}}  &
            \colhead{{$M_{\rm BH}$}}       &
            \colhead{{$K\!E_{\rm exp}$}}     &
            \colhead{{$J_{\rm final}$}}
            \\
            \colhead{[\Msun]}            &
            \colhead{}                   &
            \colhead{[$10^{52}$\,erg\,s]}   &
            \colhead{}                   &
            \colhead{[\Msun]}            &
            \colhead{[\Msun]}            &
            \colhead{[$10^{51}$\,erg]}     &
            \colhead{[\Msun]}            &
            \colhead{[$10^{51}$\,erg]}     &
            \colhead{[$10^{52}$\,erg\,s]}
            }\\
\startdata
54  &    1.2    & 1.5 & 0.0936 &  0   & 53.3 & 0.12  & 45.9 & 0.89 & 0.79 \\
56  &    1.2    & 1.5 & 0.0913 &  0   & 51.6 & 0.44  &{\bf 46.8}& 0.72 & 0.43 \\
58  &    1.2    & 1.5 & 0.0891 &  0   & 49.2 & 0.96  & 46.6 & 1.41 & 0.72 \\
60  &    1.2    & 1.5 & 0.0871 &  0   & 49.9 & 1.56  & 45.4 & 2.34 & 0.32 \\
62  &    1.2    & 1.5 & 0.0851 &  0   & 43.6 & 2.26  & 41.1 & 2.64 & 0.35 \\
54  &    1.2    &  3  & 0.0932 &  0   & 53.3 & 0.041 & 46.2 & 1.33 & 1.30 \\
56  &    1.2    &  3  & 0.0910 &  0   & 53.6 & 0.19  &{\bf 48.5}& 0.82 & 1.18 \\
58  &    1.2    &  3  & 0.0889 &  0   & 50.6 & 0.62  & 48.1 & 0.92 & 0.41 \\
60  &    1.2    &  3  & 0.0868 &  0   & 50.5 & 1.13  & 47.2 & 1.78 & 0.67 \\
62  &    1.2    &  3  & 0.0849 &  0   & 49.4 & 1.77  & 44.2 & 2.68 & 0.59 \\
56  &    1.2    &  6  & 0.0892 & 0.42 & 54.0 & 0.25  & 46.8 & 1.17 & 1.73 \\
58  &    1.2    &  6  & 0.0874 & 0.37 & 57.0 & 0.036 & 48.6 & 1.12 & 1.51 \\
60  &    1.2    &  6  & 0.0856 & 0.26 & 57.7 & 0.18  &{\bf 52.4}& 0.87 & 0.96 \\
62  &    1.2    &  6  & 0.0839 & 0.02 & 54.2 & 0.59  & 49.1 & 1.04 & 0.38 \\
64  &    1.2    &  6  & 0.0797 & 0.01 & 53.9 & 0.11  & 49.5 & 1.64 & 0.27 \\
64b &    1.2    &  6  & 0.0822 &  0   & 54.1 & 1.01  & 50.2 & 1.49 & 2.85 \\
66  &    1.2    &  6  & 0.0786 &  0   & 50.4 & 1.75  & 43.6 & 3.05 & 0.15 \\
66b &    1.2    &  6  & 0.0805 &  0   & 50.2 & 1.78  & 44.5 & 2.81 & 1.75 \\
68  &    1.2    &  6  & 0.0770 &  0   & 25.1 & 3.08  & 25.1 & 3.08 & 0.47 \\
68b &    1.2    &  6  & 0.0790 &  0   & 30.5 & 2.90  & 30.5 & 2.90 & 0.87 \\
58  &    1.2    & 12  & 0.0844 & 2.11 & 55.4 & 0.018 & 47.2 & 0.81 & 0.92 \\
60  &    1.2    & 12  & 0.0817 & 1.95 & 57.2 & 0.019 & 48.5 & 0.84 & 0.44 \\
62  &    1.2    & 12  & 0.0805 & 1.73 & 58.2 & 0.42  & 51.4 & 1.69 & 1.84 \\
64  &    1.2    & 12  & 0.0797 & 1.67 & 59.1 & 0.19  & 46.7 & 1.11 & 0.46 \\
64b &    1.2    & 12  & 0.0792 & 0.28 & 63.5 & 0.015 & 54.8 & 1.24 & 6.74 \\
66  &    1.2    & 12  & 0.0786 & 1.56 & 60.7 & 0.28  &{\bf 55.5}& 0.97 & 0.53 \\
66b &    1.2    & 12  & 0.0778 & 0.23 & 64.5 & 0.14  & 56.5 & 1.17 & 6.86 \\
68  &    1.2    & 12  & 0.0770 & 1.50 & 60.4 & 0.45  & 53.8 & 1.20 & 0.42 \\
68b &    1.2    & 12  & 0.0766 & 0.21 & 67.0 & 0.043 & 59.1 & 1.40 & 7.20 \\
70  &    1.2    & 12  & 0.0759 & 1.32 & 51.7 & 1.54  & 45.1 & 3.12 & 0.14 \\
72  &    1.2    & 12  & 0.0745 & 1.21 & 48.6 & 1.87  & 43.4 & 3.10 & 0.16 \\
74  &    1.2    & 12  & 0.0730 & 1.04 & 28.6 & 2.83  & 28.6 & 2.83 & 0.89 \\
    &           &     &        &      &      &       &      &      &      \\
62  & 1.0/1.35  &  6  & 0.177  &  0   & 60.6 & 0.058 & 58.2 & 0.28 & 2.93 \\
64  & 1.0/1.35  &  6  & 0.175  &  0   & 63.1 & 0.054 &{\bf 59.2}& 1.09 & 1.85 \\
66  & 1.0/1.35  &  6  & 0.173  &  0   & 58.6 & 0.56  & 54.5 & 1.04 & 0.82 \\
68  & 1.0/1.35  &  6  & 0.170  &  0   & 58.3 & 0.95  & 55.0 & 1.48 & 1.60 \\
70  & 1.0/1.35  &  6  & 0.168  &  0   & 54.5 & 1.35  & 50.7 & 1.99 & 1.47 \\
72  & 1.0/1.35  &  6  & 0.166  &  0   & 46.6 & 1.98  & 43.8 & 2.47 & 1.09 \\
74  & 1.0/1.35  &  6  & 0.164  &  0   &  0   & 4.84  &  0   & 4.84 &  0   \\
66  & 1.0/1.35  & 12  & 0.169  & 1.56 & 61.7 & 0.14  & 61.7 & 0.14 & 4.32 \\
68  & 1.0/1.35  & 12  & 0.167  & 1.66 & 64.2 & 0.074 & 61.3 & 0.98 & 2.19 \\
70  & 1.0/1.35  & 12  & 0.165  & 1.43 & 65.2 & 0.14  &{\bf 61.8}& 0.48 & 1.44 \\
72  & 1.0/1.35  & 12  & 0.163  & 1.36 & 64.8 & 0.29  & 61.2 & 0.87 & 0.74 \\
74  & 1.0/1.35  & 12  & 0.161  & 1.35 & 66.7 & 0.27  & 61.3 & 1.03 & 0.72 \\
76  & 1.0/1.35  & 12  & 0.159  & 0.97 & 64.5 & 0.93  & 58.0 & 2.33 & 1.68 \\
78  & 1.0/1.35  & 12  & 0.158  & 0.94 & 59.5 & 1.44  & 56.4 & 2.02 & 2.58 \\
    &           &     &        &      &      &       &      &      &      \\
68  & 0.75/1.35 &  3  & 0.244  &  0   & 68.0 &  0    &{\bf 67.9}& 0.001& 2.97 \\
70  & 0.75/1.35 &  3  & 0.241  &  0   & 67.2 & 0.10  & 67.2 & 0.10 & 2.41 \\
72  & 0.75/1.35 &  3  & 0.239  &  0   & 61.6 & 0.88  & 61.6 & 0.88 & 1.68 \\
74  & 0.75/1.35 &  3  & 0.237  &  0   & 33.7 & 2.89  & 33.7 & 2.89 & 0.36 \\
68  & 0.75/1.35 &  6  & 0.243  &  0   & 68.0 &  0    & 67.9 & 0.004& 5.87 \\
70  & 0.75/1.35 &  6  & 0.240  &  0   & 70.0 &  0    &{\bf 69.9}& 0.007& 5.92 \\
72  & 0.75/1.35 &  6  & 0.238  &  0   & 69.0 & 0.096 & 69.0 & 0.096& 4.74 \\
74  & 0.75/1.35 &  6  & 0.236  &  0   & 66.4 & 0.63  & 66.4 & 0.63 & 3.80 \\
76  & 0.75/1.35 &  6  & 0.234  &  0   & 44.1 & 2.35  & 44.1 & 2.35 & 1.19 \\
78  & 0.75/1.35 &  6  & 0.232  &  0   &   0  & 5.12  &  0   & 5.12 &   0  \\
68  & 0.75/1.35 & 12  & 0.239  & 1.39 & 66.6 &  0    & 66.3 &  0   & 8.32 \\
70  & 0.75/1.35 & 12  & 0.237  & 1.43 & 68.6 &  0    & 68.3 &  0   & 8.92 \\
72  & 0.75/1.35 & 12  & 0.235  & 1.16 & 70.8 &  0    & 69.9 & 0.033& 7.90 \\
74  & 0.75/1.35 & 12  & 0.233  & 1.19 & 70.4 & 0.12  & 70.4 & 0.12 & 6.40 \\
76  & 0.75/1.35 & 12  & 0.231  & 0.87 & 73.4 & 0.051 &{\bf 73.3}& 0.051& 8.62 \\
78  & 0.75/1.35 & 12  & 0.229  & 0.75 & 73.0 & 0.015 & 73.0 & 0.015& 7.60 \\
80  & 0.75/1.35 & 12  & 0.227  & 0.28 & 69.9 & 0.85  & 69.9 & 0.85 & 6.33 \\
80b & 0.75/1.35 & 12  & 0.227  & 0.03 & 74.9 & 0.29  & 74.9 & 0.29 & 9.11
\enddata
\tablecomments{See \Tab{helium} for the definition of $f_{\rm
    Buch}$. $J_{\rm init}$ is the initial total angular of the helium
  star. It can only decrease due to mass loss, though angular momentum
  can be transported within the star. $X(^{12}$C) is the central
  carbon mass fraction just before central carbon burning
  ignites. $M_{\rm shed}$ is the mass shed by rotation up to the point
  when oxygen burning ignites.  $M_{\rm pulse 1}$ and $K\!E_{\rm pulse
    1}$ are the mass following the first pulse and the kinetic energy
  of the initial ejecta. A zero value for $K\!E_{\rm pulse 1}$
  indicates stable oxygen ignition with no explosive mass
  loss. $M_{\rm BH}$ and $K\!E_{\rm fin}$ are the mass of the black
  hole remnant and the total kinetic energy of all ejecta including
  the first pulse. $J_{\rm fin}$ is the angular momentum of the black
  hole remnant. Models 64b, 66b, 68b, and 80b were calculated without
  including the transport of angular momentum by magnetic torques and
  are of questionable validity. Otherwise, the most massive black hole
  for each choice of reaction rates and angular momentum is given in
  boldface. Black hole masses for models with $J_{\rm final} \gtaprx
  (M_{\rm BH}/33.5 \Msun)^2$ are upper bounds. The Kerr parameter is
  the ratio of these two quantities.}  \lTab{rotate}
\end{deluxetable*}

\begin{figure}
\includegraphics[width=0.48\textwidth]{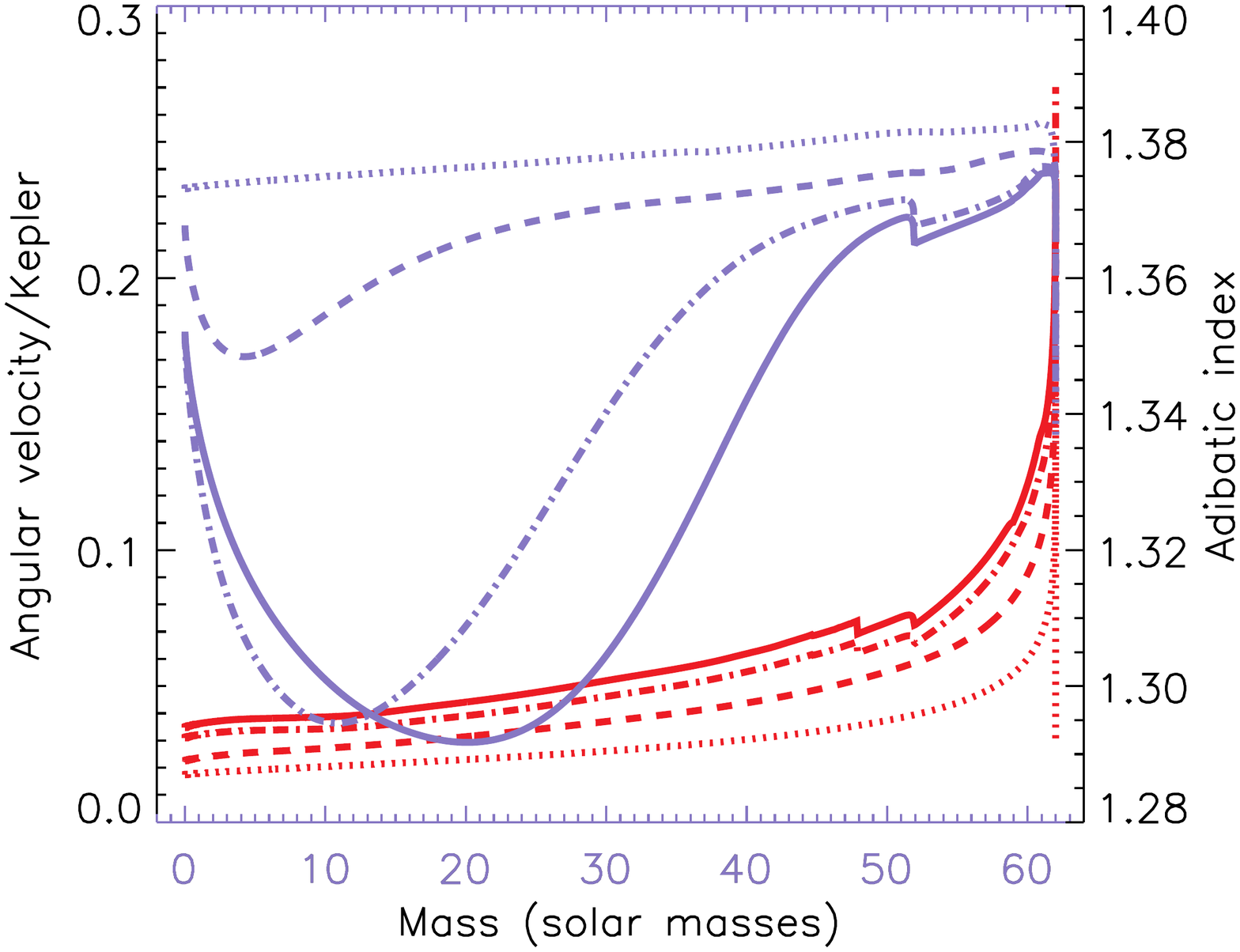}
\includegraphics[width=0.48\textwidth]{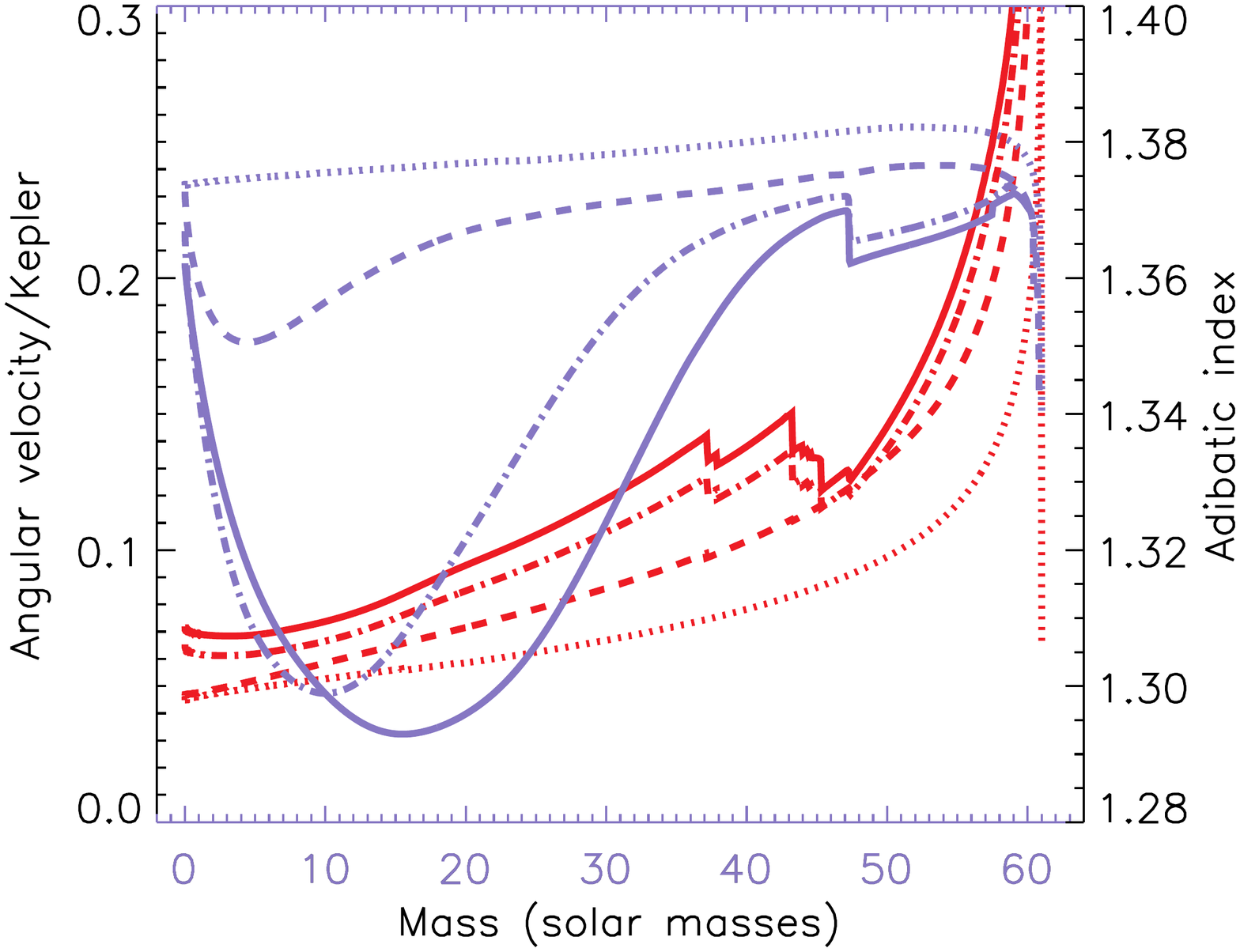}
\caption{Distribution of angular velocity and adiabatic index inside
  two 62\,\Msun\ models, Model 62$_{1.2,3}$ and Model 62$_{1.2,12}$ at
  four epochs in their evolution. The epochs are characterized by the
  first time a central temperature of $T_9 = 0.5$ (dotted line), 1.2
  (\textsl{dashed line}), 2.0 (\textsl{dash-dotted line}), or 2.5 (\textsl{solid line}) is
  encountered. These four temperatures correspond approximately to
  just prior to carbon ignition (0.5); after central carbon
  depletion (1.2); near oxygen ignition (2.0) and during the
  first pulse (2.5). The adiabatic index, $\Gamma_1$, is the blue
  lines and values less than 4/3 are locally unstable. The red lines
  are the ratios of the angular velocity in rad s$^{-1}$ to the local
  Keplerian equivalent, $(GM/r^3)^{1/2}$. The top panel is for a
  moderately rotating Model 66$_{1.2,3}$ and the bottom on for the
  more rapidly rotating Model 66$_{1,2,12}$. The inflection point at
  47\,\Msun\ for the solid red line in the lower panel is at the base
  of the helium burning shell. Global averages of these quantities
  determine the instability of the star at a given point
  (\Eqalt{instability}).  \lFig{rotcore}}
\end{figure}

\begin{figure}
\includegraphics[width=0.48\textwidth]{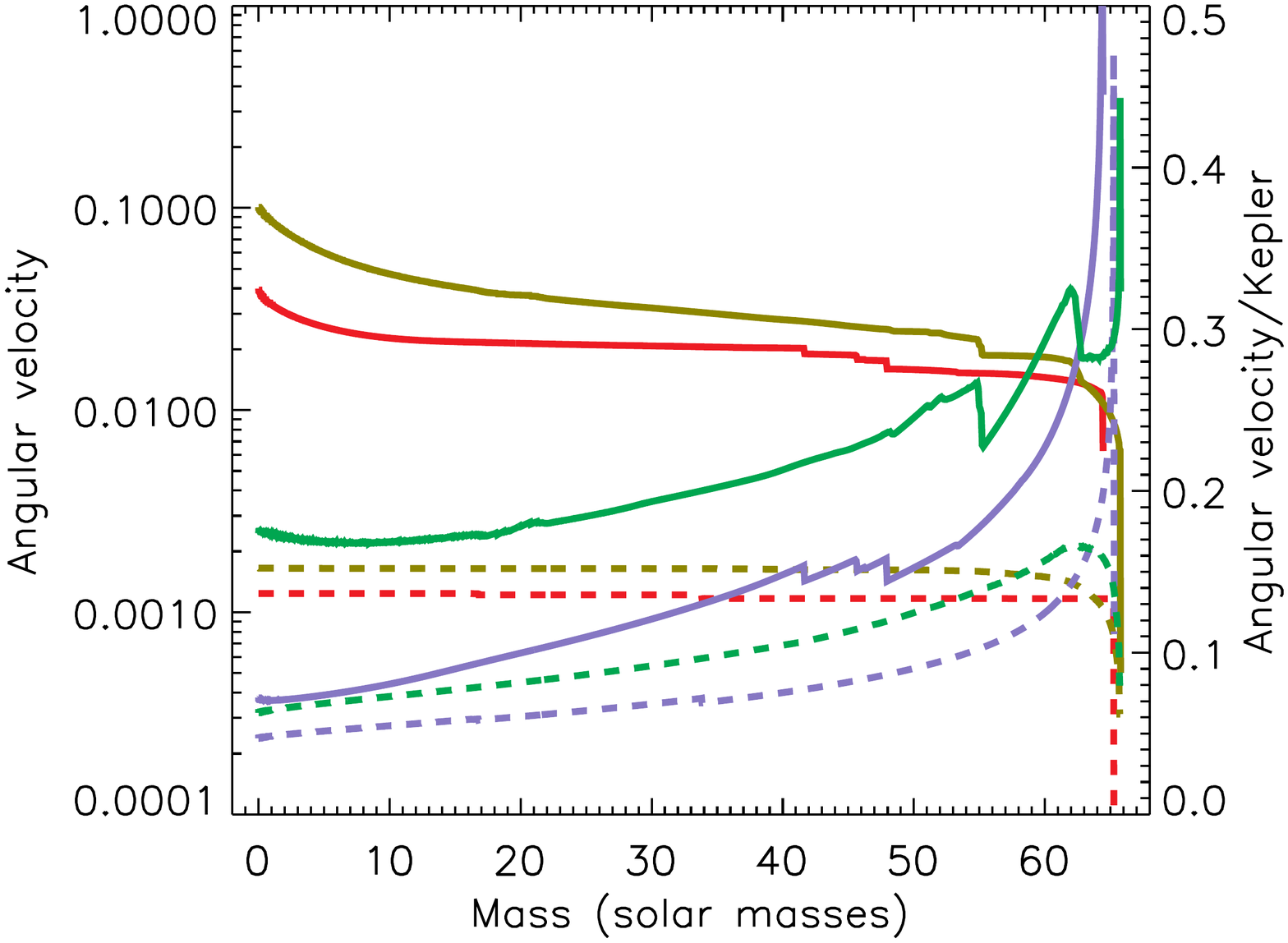}
\includegraphics[width=0.48\textwidth]{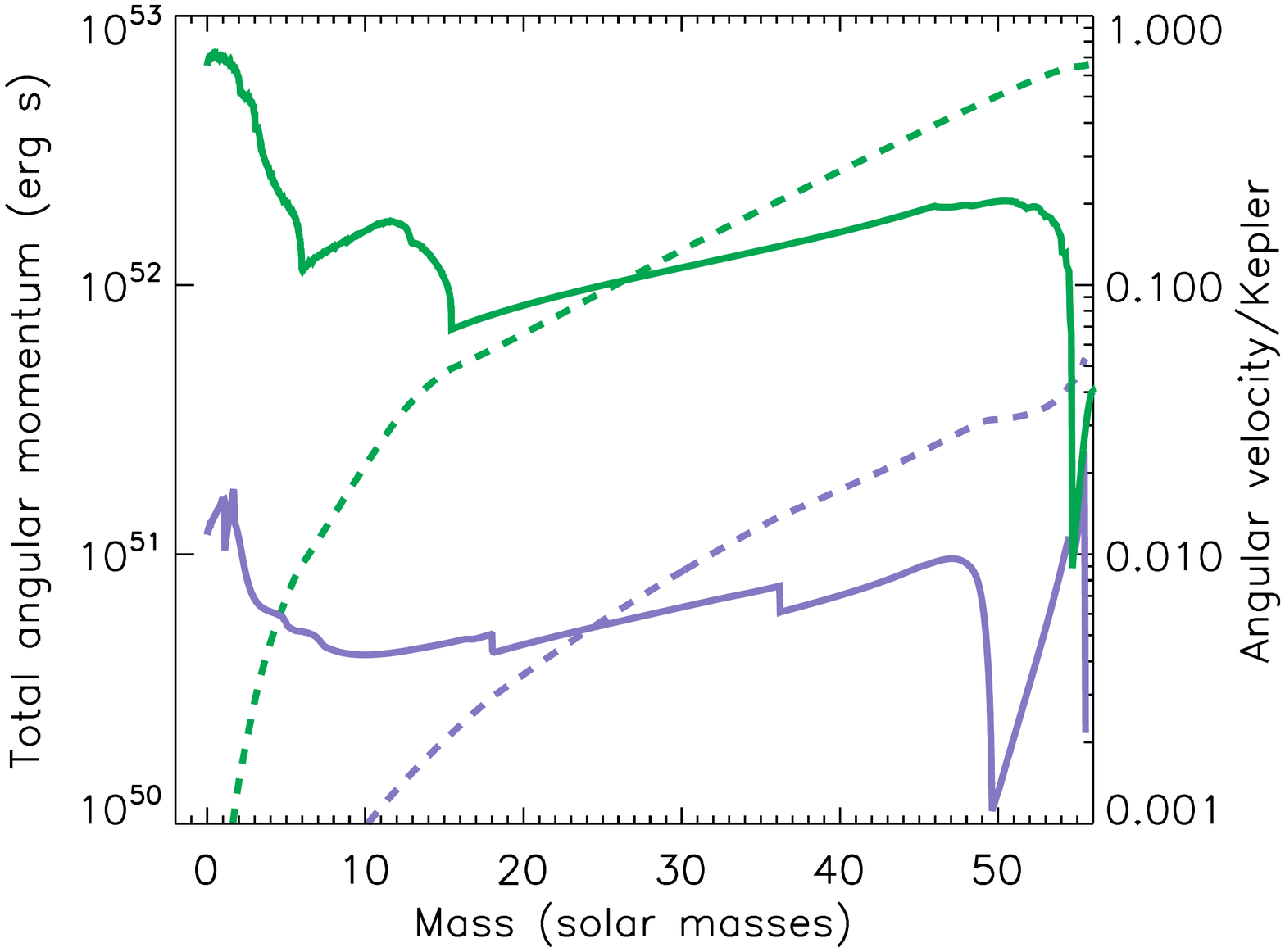}
\caption{The effect of magnetic torques on the angular momentum
  distribution in Model 66$_{1.2,12}$ (\Tab{rotate}). \textsl{Top:} The
  angular rotational speed in rad s$^{-1}$ and the ratio of angular
  velocity to Keplerian are given at two different times in the
  evolution when the central temperature is $T_9 = 0.5$ (just prior
  to carbon ignition; dashed lines) and 2.5 (during the first
  pulse; solid lines).  Blue lines are the rotational velocity for
  cases where magnetic torques are included in the calculation; green
  lines are for an equivalent model without magnetic torques.  Red
  lines give the ratio of angular speed to Keplerian for the model
  with magnetic torques; gold lines are for the model without
  torques. \textsl{Bottom:} The continued evolution of the star leads to even
  greater disparity between the magnetic and non-magnetic cases as is
  shown here for the presupernova star. The total angular momentum
  interior to the given mass and the ratio of the local rotation rate
  to Keplerian are given as dashed and solid lines respectively. The
  iron core of the non-magnetic case would experience triaxial
  deformation in a multi-dimensional study.  \lFig{nob}}
\end{figure}

Prior to any pulsational mass loss, all models with $J_{\rm init}$ =
$1.2 \times 10^{53}$\,erg\,s and many with $J_{\rm init} = 6 \times
10^{52}$\,erg\,s experience rotational mass shedding. This begins late
in helium burning, but mostly happens near and just after carbon
ignition. The mass lost is $\ltaprx2$\,\Msun, which is a small
fraction of the total mass, but often contains a substantial
fraction of the total angular momentum. For example, Model
60$_{1.2,12}$ shrank to 59.04\,\Msun\ at carbon ignition and, in the
process, its angular momentum, $J_{52}$, declined to from 12 to
7.84. By oxygen ignition, these values had further decreased to
58.05\,\Msun\ and 6.35, respectively. Even before pulsations began,
the initial angular momentum for Model 60$_{1.2,12}$ had almost
halved, making Model $58_{1.2,6}$ and Model $60_{1.2,12}$ similar in
outcome. The two stars had masses of 57.6 and 58.0 \Msun \ after
rotational mass shedding and produced black holes of 48.6\,\Msun\ and
48.5\,\Msun. Other pairs, like 62$_{1.2,6}$ and 64$_{1.2,12}$ did not
yield such similar results because of the stochastic nature of
multiple pulses and  sensitivity to slight changes in the angular
momentum distribution and composition. The agreement was especially
poor for models near the threshold of full explosion like 66$_{1.2,6}$
and $68_{1.2,12}$, and for models with little rotational mass shedding.

Still, the implication is that it would not be fruitful to consider
models with much larger rotation rates than $J_{\rm init,52}=12$. The
excess momentum would simply be shed early in the evolution and the
maximum black hole mass would not be greatly increased. This is
particularly true since the criterion for rotational mass loss,
$\omega_{\rm k}=0.5$ at the surface, is already too large for
stars so near the Eddington limit.

Also shown in \Tab{rotate} is the energy of the first pulse and the
mass of the star after that pulse (including the effects of rotational
mass shedding prior to the pulse). These characteristics follow
expectations more consistently than final properies. For a given
angular momentum, the mass ejected and the kinetic energy of that
mass, $K\!E_{\rm pulse1}$ increase with initial mass. For a given
initial mass, $M_{\rm pulse1}$ increases with initial angular
momentum, while $K\!E_{\rm pulse1}$ decreases. In cases where the
central carbon mass fraction and initial angular momentum are high,
the stars often ignite central oxygen burning stably. Such stas may
also experience a long phase of numerous, weak pulsations due to the
burning of carbon in a shell. The mass lost prior to oxygen burning in
such models is small though.

A strong first pulse on the other hand implies a long wait before the
next pulse while the star relaxes from a highly extended, weakly bound
state. During this time, angular momentum can be transported out of
the core by shear, magnetic torques, and convection. For these cases,
subsequent pulses are not so influenced by rotation as the first one
and can also be quite strong.

Indeed, the transport of angular momentum by magnetic torques prior to
oxygen ignition and during the long interpulse periods, when they
exist, is a major uncertainty in these calculations. Eliminating them
reduces rotational mass shedding. Angular momentum that might have
been transported to the surface and lost remains concentrated in the
central core where it influences the subsequent evolution. \Fig{nob}
shows the evolution of the angular velocity and the ratio of the
angular velocity to Keplerian, $\omega_{\rm k}$, for Model
66$_{1.2,12}$ with and without magnetic torques (models in
\Tab{rotate} that end in b, like ``66b'' did not include magnetic
torques). Early on, near carbon ignition at $T_9 = 0.5$, the
rotational profiles are similar for the two models, though rotational
mass shedding has already reduced the angular momentum in the model
without magnetic fields. As time passes, the divergence increases and
by the time the iron core collapses, there is more than an order of
magnitude difference in the angular momenta of the remnants.

The chief cause of divergence in this case is magnetic braking during
the thousand year long interval following the second pulse in the
model that includes magnetic fields. During that time, the angular
momentum in the inner 55.5\,\Msun\ of the star (the part destined to
be a black hole) declines from $4.2 \times 10^{52}$\,erg\,s to $5.4
\times 10^{51}$\,erg\,s.  The mass of the star is constant during this
interval so the excess angular momentum piles up in a mass of just
a few {\Msun}\ near the surface. This is later ejected in a final third
pulse so that the star that included magnetic fields ends up making a
black hole with spin $J_{52} = 0.53$ whereas the one without magnetic
fields makes a black hole of nearly the same mass, but with $J_{52} =
6.86$.

It would not be realistic to ignore magnetic torques during these long
Kelvin-Helmholtz episodes while cores recover from prior explosive
expansion. The compositions, densities. temperatures and rotation
rates are similar to those in stable carbon burning. On the other
hand, experimental validation of the simple prescription used is
lacking. The correct solution might lie between the two extremes. This
means that any estimate of the Kerr parameter for the black holes
coming from these rapidly rotating models, especially those with large
carbon abundance is not reliable.


The dimensionless Kerr Parameter is $a/M = cJ/GM^2 = J_{52}
(33.5\,\Msun/M)^2$.  For a 70\,\Msun\ black hole with $a/M = 1$, the
critical angular momentum is $4.40 \times 10^{52}$\,erg\,s. Compact
remnants in \Tab{rotate} with much more angular momentum than this
cannot form black holes without losing mass and shedding the
excess. This includes many of the more massive models with large
carbon mass fractions in \Tab{rotate}. Only a small fraction of this
angular momentum is in the collapsing iron core itself. Most is in the
outer layers. The most likely fate of such cores is to form a black
hole surrounded by an accretion disk. This probably leads to the
production of jets \citep{Woo93,Mac99} and mass ejection. That is,
these are excellent candidates for making long soft gamma-ray bursts, and
GRBs may be intimately related to the formation of the most massive
LIGO sources \citep[see also][]{Mar20}.  The bursts, of course,
happened long ago when the black holes formed, not when they
merged. They are probably not the most common form of GRBs since they
involve quite massive black holes and the surface layers of the
star. Given the large angular momentum, the disk might form at large
radius and thus have a long characteristic time scale.  A time scale
of minutes is implied by the free fall time for the outer layers and
not an unreasonable estimate for the burst duration.

How much mass is ejected? Most of the final angular momentum is in the
outer part of the remnant (\Fig{nob}). Consider the case of
76$_{0.75/1.35,12}$. This model leaves a 73.3\,\Msun\ remnant with
$J_{\rm final,52} = 8.62$, too much angular momentum for a black hole
with this mass. But $J_{52} = 4.2$ exists at 69\,\Msun \ corresponding
to $a/M = 1$. A probable outcome then is that a black hole with an
initial mass $\sim$69\,\Msun\ forms with an accretion disk of about
4\,\Msun. Jet formation will eject some, but not all of this 4\,\Msun.
Based on calculations by \citet{Zha04}, the fraction ejected might not
be large. It depends on the efficiency with which lateral shocks
induced by the jet eject matter in the equatorial plane.  The
impication is that some of the large black hole masses in \Tab{rotate}
should be treated as upper bounds, but are perhaps not too far from
correct. The final black hole of all models with $J_{52} \gtaprx
(M/33.5\,\Msun)^2$ would have a dimensionless Kerr parameter near
$a/M=1$.

The maximum mass black hole for a given choice of reaction rates and
angular momentum is given in boldface in \Tab{rotate}. Models without
magnetic torques are ignored. These limits range from little
modification to the those derived for non-rotating stars in
\Tab{helium} if $J_{\rm init,52} \ltaprx 3$, to increases as great as
11\,\% for maximal rotation.  Our heaviest black hole is 73\,\Msun
\ (though, as previously noted, that model has too much spin to fully
collapse to a black hole). This is a larger increase than estimated by
\citet{Mar20} for stars with magnetic torques included. The
difference probably involves the consideration here of more rapid
rotation and, especially, choices of reaction rates that produce more
carbon. Without magnetic fields the increase could potentially be
greater, though probably not much larger as Model 80b illustrates.









\subsection{Progenitor Evolution}
\lSect{prog}

The rotation rates assumed here are similar to those used to model GRB
progenitors \citep{Woo06,Yoo05,Yoo06,Can07}. Chemically homogeneous
evolution is common for these stars, and is consistent with the need to
keep LIGO progenitors compact inside close binaries \citep{Man16}.

Consider a 70\,\Msun\ main sequence star with initial angular momentum
$J = 4 \times 10^{53}$\,erg\,s. This corresponds to an equatorial
rotational speed at the surface of 390\,km\,s$^{-1}$ and an angular
velocity there that is 32\,\% Keplerian. With this rotation rate, the
star evolves chemically homogeneously \citep{Man16}. It never becomes
a giant and consumes all the hydrogen within its mass, except for a
small fraction near the surface, while still on the main sequence. It
thus evolves directly into a helium star similar to the ones
considered in this section.

If mass loss were negligible, this initial angular momentum
would be preserved and the star would die with values in excess of
the largest ones in \Tab{rotate}, but that is not realistic. On
the other hand, with solar metallicity and conventional mass loss
rates, the star would end up with too small a mass to encounter the pair
instability and would have a negligible amount of angular momentum.

Consider then the results of substantially decreasing the star's mass
loss rate, perhaps due to a small initial metallicity. Following
\citet{Sze15}, we consider a metallicity of 2\,\% solar, more
specifically an iron mass fraction of $2.92 \times 10^{-5}$. Their
67\,\Msun\ model had a rotational velocity of 400\,km\,s$^{-1}$,
comparable to the 390\,km\,s$^{-1}$ for our 70\,\Msun\ model. Their
model also evolved chemically homogeneously. For the mass loss
prescription they employed, their star had a mass at the end of
hydrogen burning of 62.2\,\Msun. Most of their mass loss occurred
after the star became a Wolf-Rayet star with a surface hydrogen mass
fraction less than 0.3.

\citet{Sze15} used a complicated prescription for mass loss involving
multiple sources and interpolation. Here we just use the rates of
\citet{Nie90} and \citet{Yoo17} with a multiplicative factor less than
one for the Yoon rate.  Given the uncertain weak dependence of Yoon's
mass loss rate on metallicity, these reductions are reasonable.
With this prescription, our 70\,\Msun\ metal-poor star reaches a
central hydrogen mass fraction of 0.3 with a mass of 69.4 \Msun. Up to
that point mass loss was taken from \citet{Nie90} and was
small. Continued evolution used the Yoon rates with a multiplicative
factor 0.5. The star finished hydrogen burning with a mass of
63.7\,\Msun\ and a residual angular momentum of $2.3 \times
10^{53}$\,erg\,s. The model of \citet{Sze15} had a luminosity at
hydrogen depletion of 10$^{6.34}$\,\Lsun\ and an effective temperature
of 80,600 K.  Our values were $10^{6.37}$\,\Lsun\ and 86,800\,K. The
slight mismatch in starting mass and rotation rate is inconsequential
compared with uncertainties in the mass loss rate.

It is necessary to follow the continued evolution through helium
burning. Published mass loss rates for metal-poor helium stars span a
wide range \citep{Vin17,Yoo17,San19,San20,Woo20}. Here we continued to
use the rate of \citet{Yoo17} scaled, for the given metallicity (2\,\%
solar), by factors of 1, 0.5, and 0.25. All runs were started from the
model with central hydrogen mass fraction 0.3.  These multipliers gave
a range of {\sl average} mass loss rates during helium burning from $4
\times 10^{-5}$\,\Msun\,y$^{-1}$ to
$1\times10^{-5}$\,\Msun\,y$^{-1}$. The instantaneous rate was about a
factor of two smaller early in helium burning when the star was still
a WNE star and larger later on when it is a WC star.  These rates are
consistent with what \citet{San20} have recently published for a
metallicity of 2\,\% to 5\,\% solar. With these three rates, the final
mass of the WC stars at carbon ignition were 47.2, 56.7, and
62.7\,\Msun\ and their angular momenta were 2.7, 9.7, and $19\times
10^{52}$\,erg\,s. This spans most of the range of angular momenta
sampled in \Tab{rotate}. Other choices of initial mass and angular
momentum could have filled the grid.


\section{Super-Eddington Accretion}
\lSect{accrete}

Another source of uncertainty affecting black hole masses in a close
binary system is the subsequent interaction of the first black hole
born with the other star \citep{Van20}.  It is well known that a black
hole can accrete at a rate well above the Eddington limit. Accretion
rates of solar masses per second are not uncommon during core collapse
simulations in failed supernovae.  The excess energy that might have
inhibited infall or powered mass outflow is either radiated away as
neutrinos or advected into the event horizon and lost \citep{Pop99}. A
less extreme and more relevant example is Model r003 of
\citet{Sad16}. In their 3D general relativistic MHD simulation, a 10\,\Msun\ black hole accretes at 176 times the Eddington value, that is
at at rate $6.7\times 10^{-6}$\,\Msun\,y$^{-1}$. The authors note
that the calculations should scale for larger mass black holes and
that greater super-Eddington factors are allowed. For an 80\,\Msun
\ black hole, Model r003 would correspond to an accretion rate of
about $5 \times 10^{-5}$\,\Msun\,y$^{-1}$. We speculate that similar
flows and characteristics would characterize accretion rates on up to
0.001\,\Msun\,y$^{-1}$. \citet{Sad16} note that a portion of the
accreted energy, about 3\,\% $\dot M c^2$, would go into powering
semi-relativistic polar outflows.  Black holes undergoing this sort of
accretion might thus appear as ultra-luminous x-ray sources or SS433
analogues. Indeed, an accretion rate of 10$^{-4}$\,\Msun\,y$^{-1}$ has
been inferred for SS 433 \citep{Che20}. Hyper-Eddington accretion
factors of up to 1,000 have been computed for supermassive black holes
including the effects of the outgoing jet on the disk \citep{Tak20}.

It is beyond the scope of this paper to consider the full complexity
of a mass exchanging binary, especially those that might experience
common envelope evolution, but our single star models offer some
insights.  The progenitors of the more massive LIGO sources were
probably stars of $\sim150$\,\Msun\ or more and possibly formed in
metal deficient regions. One or more stages of common envelope
evolution are frequently invoked. The accretion by a black hole during
a dynamic common envelope phase is currently thought to be small
\citep{De20}, but the evolution leading up to that common envelope is
poorly explored for stars of such great masses.

We consider here a 130\,\Msun\ star finishing its hydrogen burning
evolution in a close binary with a black hole of $\sim60$\,\Msun\ with
an initial orbital separation of 300\,\Rsun, or about $2 \times
10^{13}$\,cm. Such a configuration is reasonable for the progenitors
of LIGO sources \citep{Bog18}. The initial black hole mass is
reasonable for the range of reaction rates and rotation rates
considered in this paper. We further assume that the star has 10\,\%
solar metallicity. Its mass after hydrogen burning is thus about
116\,\Msun\ (\citealt{Woo17}; Model T130).  In the initial
configuration the period of the binary is 40 days. The radius of the
main sequence star is $\sim 1.5 - 2$\,\Rsun\ throughout hydrogen
burning. The main sequence star does not yet fill its Roche lobe.

The situation changes as the star completes hydrogen burning,
contracts in its center to ignite helium burning, and attempts to
become a supergiant. The star develops a surface convective zone that
gradually eats into its hydrogen envelope.  The rate at which the
convective zone moves into the envelope, which is also the time to
transition to a supergiant, depends on the theory of semi-convection,
but is of order a nuclear time scale, or $\sim10^5$\,yr for helium
burning. The Kelvin-Helmholtz time scale for the envelope is smaller,
a few thousand years.  If the envelope were to be lost on either time
scale, the loss rate would thus correspond to 0.0002 to
0.002\,\Msun\,y$^{-1}$. Until close to the end, most of the envelope,
by mass, remains dense transports radiation by diffusion except for a
region just above the hydrogen burning shell.

We explored the effect of imposing a limit on the radius of this
116\,\Msun\ secondary at $2 \times 10^{13}$\,cm, i.e., manually
removing all mass beyond this radius as the star attempted to become a
giant. A steady state was found with a surface convective zone that
had an outer boundary that extended nearly to the cut-off radius. The
base of this convective shell remained fixed at $\sim7 \times
10^{12}$\,cm as mass was removed from the top and new matter
convectively dredged up from the bottom.  At any time, this convective
shell contained only a few hundredths of a solar mass with density
ranging from 10$^{-11}$\,g\,cm$^{-3}$ at the surface to
10$^{-8}$\,g\,cm$^{-1}$ at its base. The binding energy of this
convective shell, $\sim10^{46}$\,erg, was very small compared with the
orbital kinetic energy of the black hole, $\sim10^{50}$\,erg. The
Roche radius was inside of this convective envelope, close to its
base, but the angular velocity of the black hole within this envelope
was supersonic, so stable Roche lobe overflow seems unlikely.

Rather we speculate that the friction resulting from the black holes
motion drives mass loss from the secondary. \citet{Pod01} has
discussed such ``frictionally driven winds''.  If the convective layer
is removed, the friction is reduced so the mass loss might be self
regulated and determined by the rate at which convection moves into
the envelope. We followed our toy model for 100 years (about 10,000
steps) , during which the mass loss and convective shell properties
remained in steady state. The inferred mass loss rate was
0.0018\,\Msun\,y$^{-1}$.

This is a common envelope evolution of sorts, but one that proceeds on
a thermal time scale, not a dynamical one \citep{Pod01}. The
efficiency for the black hole to retain the mass that is lost by the
secondary is unknown, but a total envelope mass of 60\,\Msun\ could be
lost this way. Growth of the hole by 15\,\Msun\ or more (half of the
envelope lost to a wind and half that accreted) does not seem
unreasonable \citep[see also][]{Van20}.  Over time, the black hole
gains mass and, initially at least, moves closer to the secondary. It
could be that eventually the interaction becomes stronger as the orbit
is narrowed and a more dynamical sort of common envelope evolution
occurs. The 60\,\Msun\ helium core of the secondary eventually
experiences its own pair instability, making a black hole not much
smaller than 60\,\Msun\ and, if the black holes have become close
enough, owing to the action of common envelope evolution jets, or
kicks, they merge on a Hubble time. This is clearly a very uncertain
and possibly quite rare model.

\section{Discussion}
\lSect{conclude}

Four modifications to the assumptions previously used to calculate the
``pair-gap'' in the black hole birth function have been explored. Each
has been suggested previously and has the potential to substantially
alter the boundaries, but each also has its problems. In general, it
is not too difficult to raise the lower bound of the cut off, \Mlo,
from 46\,\Msun\ \citep{Woo19} to about 60 or even 70\,\Msun, but
raising it to 85\,\Msun\ and more is difficult and might require
either that the very massive black hole was captured, i.e., not born
in the binary that subsequently merged, or other special
circumstances. That is, pending better determination of uncertain
reaction rates, the events in the GWTC-2 catalog \citep{Abb20c} might be
accounted for with rather standard physics in which there is still a
pair gap with $\Mlo \sim 60$\,\Msun, but GW190251 requires an
alternate explanation. Two possible explanations were discussed here -
birth in a detached binary (\Sect{single}) and super-Eddington
accretion after black hole birth (\Sect{accrete}). See also
\citet{Tut17} and \citet{Van20}.

The combination of $^{12}$C($\alpha,\gamma)^{16}$O and $3 \alpha$
rates used in our previous studies may have conspired to produce a
small value for \Mlo.  Using the favored value for
$^{12}$C($\alpha,\gamma)^{16}$O from \citet{Deb17}, 140\,keV\,b for the
$S$-factor, gives 48\,\Msun\ (\Tab{helium} with $f_{\rm Buch} = 1.0$
corresponds to $S = 146$\,keV\,b). If one accepts a major revision to the
$3 \alpha$ rate \citep{Kib20,Eri20} but does not change
$^{12}$C($\alpha,\gamma)^{16}$O, \Mlo rises to 57\,\Msun.
Using the one-sigma lower bound from \citet{Deb17} for the
$^{12}$C($\alpha,\gamma)^{16}$O $S$-factor (i.e., $S = 110$\,keV\,b) and the
revised $3 \alpha$ rate from \citet{Kib20} gives $\Mlo = 64$\,\Msun, which we presently regard as an upper limit on the likely
effects of rate modification alone. This could be augmented by the
effects of rotation and future accretion by the black hole, but the
direct collapse of an 85\,\Msun\ Wolf-Rayet star to a black hole
\citep{Bel20} seems unlikely. Rate determinations are best left to the
nuclear laboratory, but using this large value of $3 \alpha$ and a
reduced value for $^{12}$C($\alpha,\gamma)^{16}$O may lead to
difficulties in stellar nucleosynthesis that have yet to be fully
explored (\Sect{nucleo}).

A larger range of reaction rates was considered by \citet{Far20}, but
within the range mutually considered, our results agree reasonably well
with \citet{Far19} who gave an upper bound on \Mlo of 56\,\Msun.  Due to sparse sampling and the cumulative effects of
  multiple pulses and burning shells, and based on examining hundreds
  of models (not all included in the tables), all values given here
  for \Mlo probably have an uncertainty of about $\pm2$\,\Msun. A value of $\Mlo = 45 - 65$\,\Msun\ is reasonable for
  now given our current understanding of the nuclear physics.

Similar rate modifications also imply a significant shift in the upper
boundary for the pair gap, \Mhi with values ranging from 136 to
161\,\Msun. Even for our standard rates, $f_{\rm Buch} = 1.2$ and $f_{3
  \alpha} = 1$, there is a significant shift of \Mhi from
133.3\,\Msun\ \citep{Heg02} to 139\,\Msun\ because of a better treatment of
the nuclear burning and a different treatment of convection during the
implosion (\Sect{mhi}).

As has been known for some time, single stars that retain part of
their hydrogen envelopes when they die can produce more massive black
holes. The previous approximate limit for this mass was 65\,\Msun
\ \citep{Woo17}. Here we have shown that the same adjustments to
reaction rates that raise \Mlo and \Mhi for close binaries will also
raise the limit for single stars. Our estimate for the maximum mass
allowed by variation of reaction rates within their current errors is
about 90\,\Msun\ (\Tab{single}). This has several implications. If one
can make an 85\,\Msun\ black hole in an isolated star and later
capture it ina binary, then there only needs to be one stage of
capture, not two as would be required if the limit were
65\,\Msun\ (one to make the 85\,\Msun\ black hole and another to make
the final system). Second, the 85\,\Msun\ black hole might be made in
a detached binary, especially in a blue supergiant or luminous blue
variable. If the secondary later became a red supergiant because of
primary nitrogen production, the black hole and secondary might still
experience a common envelope phase and draw closer. This might be a
rare event.

Depending a number of uncertainties, rotation could cause moderate to
major modifications to \Mlo (\Sect{rotation}).  The maximum
effect for our models that include magnetic torques are larger, by
about a factor of two, than those of \citet{Mar20}, probably because
we considered a larger range of reaction rates and angular
momenta. Rapidly rotating stars that experience chemically homogeneous
evolution and lose less than 30\,\% of their mass during hydrogen and
helium burning can end their lives with total angular momenta like
those in \Tab{rotate}. The resulting black hole masses will be
appreciably increased beyond those of non-rotating stars and the lower
bound on the pair-instability mass gap could be as large as
$\sim70$\,\Msun. Uncertainties include:

\begin{itemize}

\item The carbon abundance at the end of helium burning, hence the
  rates for $^{12}$C($\alpha,\gamma)^{16}$O and $3 \alpha$.

\item The initial angular momentum and the mass loss. The initial
  angular momentum for the helium star needs to be $\gtaprx 3 \times
  10^{52}$\,erg\,s to have a major effect. This is reasonable for
  rapidly rotating stars experiencing chemically homogeneous
  evolution with low mass loss rates(\Sect{prog}). Using current
  estimates in the literature for mass loss rates, the maximum
  metallicity would be between about 1\,\% and 5\,\% solar.

\item Rotational mass shedding. For an angular momenta in excess of $6
  \times 10^{52}$\,erg\,s, much of the excess can be shed before the
  pair instability is encountered.

\item Magnetic torques. These are particularly important and uncertain
  during the long interpulse periods for events that have a strong
  first pulse. Too many GRBs and rapidly rotating pulsars would exist
  if they were left out altogether, but their magnitude is uncertain.

\item Disk formation during the collapse. For those models with high
  carbon abundance, rapid initial rotation, or weak magnetic torques,
  there may be too much angular momentum for the entire presupernova
  star to collapse to a black hole. Disk formation and the
  accompanying disk winds and jet-induced explosions may shed the
  excess angular momentum, but the amount of mass ejected is
  uncertain. It might be small. If rotation is responsible for the
  most massive black holes, the final dimensionless Kerr parameter could be close to  1.

\end{itemize}

The effects of rotation on \Mhi were not explored but might be
smaller owing both to the stronger pair instability in such massive stars
and the likelihood of rotational mass shedding in stars so near the
Eddington limit.

Perhaps the most uncertain aspect of \Mlo is the accretion
the black hole experiences after it is born. This might occur either
during the common envelope phase that is frequently invoked to bring
the stars together or during frictionally-induced mass loss by the
secondary after it finishes hydrogen burning. Here we focused on the
latter. As the secondary first expands beyond about 10$^{13}$\,cm the
surface convective zone contains little mass. Most of the envelope is
radiative. If that small mass spills over onto the black hole, it will
be replenished on a thermal time scale by the envelope. Envelope
masses may be $\sim50$\,\Msun\ and thermal time scales are thousands
of years so accretion rates of $10^{-3} - 10^{-2}$\,\Msun\,y$^{-1}$
might occur. At such grossly super-Eddington rates the black hole will
accrete most of the matter, but also produce strong bipolar outflows
\citep{Sad16}.

\section{Acknowledgments}

This work has been partly supported by NASA NNX14AH34G and by the
Australian Research Council (ARC) Centre of Excellence (CoE) for
Gravitational Wave Discovery (OzGrav), through project number
CE170100004.  AH has been supported, in part, by the National Science
Foundation under Grant No. PHY-1430152 (JINA Center for the Evolution
of the Elements); and by the Australian Research Council Centre of
Excellence for All Sky Astrophysics in 3 Dimensions (ASTRO 3D),
through project number CE170100013.

\end{document}